\documentclass[acmtog,preprint]{acmart}
\renewcommand\footnotetextcopyrightpermission[1]{}
\usepackage[utf8]{inputenc}
\pdfoutput=1
\renewcommand\footnotetextcopyrightpermission[1]{} % removes footnote with conference info
\setcopyright{none}
\usepackage{hyperref}
\hypersetup{pdfstartview=FitH,pdfpagelayout=SinglePage}
\usepackage{graphicx, xspace, outlines,url,grffile,enumerate,listings,appendix,algorithm}
\usepackage[normalem]{ulem}
\usepackage{units}
\usepackage{xcolor}
\usepackage[font={footnotesize,it}]{caption}
\usepackage{paralist}
\usepackage{amsthm}
\usepackage{amsmath}
\usepackage{amsfonts}
\usepackage{microtype}
\usepackage{siunitx}
\usepackage{multirow}
\usepackage{outlines}
\usepackage[small,compact]{titlesec}
\usepackage{enumitem}
\usepackage{booktabs}
\setitemize{itemsep=1pt,topsep=1pt,parsep=1pt,partopsep=1pt}
\newcommand{\bi}{\begin{itemize}}
\newcommand{\ei}{\end{itemize}}

\newcommand{\eg}{{\it e.g.,}\xspace}
\newcommand{\ie}{{\it i.e.,}\xspace}
\newcommand\eat[1]{}

\newcommand{\allnotes}[1]{}
\renewcommand{\allnotes}[1]{\textit{#1}}

\usepackage{array}
\usepackage{ragged2e}
\usepackage[printwatermark]{xwatermark}
\newcolumntype{P}[1]{>{\RaggedLeft\hspace{0pt}}p{#1}}
\usepackage{enumitem}
\setitemize{noitemsep,topsep=0pt,parsep=0pt,partopsep=0pt}
\makeatletter
\let\@authorsaddresses\@empty
\makeatother
\makeatletter
\def\blfootnote{\gdef\@thefnmark{}\@footnotetext}
\makeatother
\begin{document}
\pagenumbering{arabic}
\begin{abstract}
\section{Abstract}
\noindent Deep Neural Nets have hit quite a crest\\
But physical networks are where they must rest\\
And here we put them all to the test\\
To see which network optimization is best
\end{abstract}
\title{How to Train your DNN: \textit{The Network Operator Edition}}
\author{Michael Alan Chang, Domenic Bottini, Lisa Jian, Pranay Kumar, Aurojit Panda, Scott Shenker}
\maketitle
\thispagestyle{empty}
\pagestyle{empty}
\section{Introduction}

Deep Neural Networks have gained significant traction in both academia and industry. One type of deep neural network -- feedforward convolutional neural networks (CNNs) -- have been the driving thrust behind critical applications such as image recognition ~\cite{DBLP:journals/corr/SimonyanZ14a, DBLP:journals/corr/HeZRS15}, drug discovery ~\cite{DBLP:journals/corr/WallachDH15}, and medical diagnosis ~\cite{DBLP:journals/corr/WallachDH15,Tomov2018OnDN, LITJENS201760, 8241753}. The accuracy of CNNs often requires frequent retraining, so it is important to reduce CNN training time. Efforts to do so have targeted nearly all layers of the software and hardware stack, and increasingly involves distributing training across machines in a cluster. Early work on distributed CNN training adopted the \emph{parameter server model}~\cite{Li2014ScalingDM} where computation is performed by several worker nodes and one or more \emph{parameter servers} are used to aggregate and distribute results from individual workers. Recent work has also looked at a variety of topics, such as improving the performance of distributed CNN training through the use of better scheduling~\cite{DBLP:journals/corr/abs-1803-03288}, and improving network transfers~\cite{Vishnu2016DistributedTW}. In this paper we focus on optimizations that involve the cluster network. A variety of network oriented solutions have been proposed, but with little resulting clarity about which proposal (or combination of proposals) achieves the best end-to-end performance. This paper seeks to answer this question.

We first observe that one can divide network optimizations into two broad categories: those that change the network fabric and those that only change software at end hosts. This is a useful distinction because changing the network fabric is typically more difficult (involving router software and perhaps hardware), while host software changes are significantly easier to implement. The first fabric-based change we consider is using IP multicast (a feature that is supported by most routers but often not enabled). IP multicast has been used previously to accelerate HPC workloads using MPI~\cite{Chen2000MPICO, Yuan2002GroupMS, Hoefler2007APC}, which in turn has been used as a communication primitive by a number of distributed CNN frameworks~\cite{Vishnu2016DistributedTW, Awan2018ScalableDD}. The other class of fabric-based optimization is in-network aggregation, as can be implemented using programmable switches. The use of in-network aggregation for CNNs has already been proposed in Daiet~\cite{Sapio2017InNetworkCI} and Luo~\cite{LuoKrishnamurthy}, and here our goal is to understand its performance impact relative to other network optimizations.

The host-based techniques we consider move away from the parameter server model. These include ring-reduce~\cite{baiduallred} and all-reduce (\eg Rabenseifner~\cite{10.1007/978-3-540-24685-5_1} or butterfly mixing~\cite{DBLP:conf/sdm/CannyZ13}) which avoid the use of a centralized server for aggregation and have been successfully used to speed up HPC jobs in the past. We provide a more detailed list of such approaches in \S\ref{all_reduce_background}. 

We analyze these various approaches -- in isolation and in combination where possible -- to answer four questions. First, how do these various optimizations rank in terms of effectiveness? Second, given this ranking, is it necessary for us to resort to fabric-based mechanisms, or do host-based mechanisms suffice? Third, how robust are these results to possible future changes in CNNs (\eg more layers)? Fourth, how robust are these results to possible future changes in host behavior (\eg changes in TensorFlow)?

We rely on trace driven simulation to address these questions. The use of simulations allowed us great latitude in testing a variety of proposals including ones which require changes to the network hardware and were hence infeasible for us to test in practice. The use of a simulation also allowed us to avoid certain approximations and non-determinism that would have been difficult to consider in a tractable analytical model. Thus, simulation provided us with a good balance between the accuracy of our results and the ability to try out a wide range of optimizations. To further ensure realism for our results we seed our simulations with traces generated from training CNNs on real hardware using distributed TensorFlow. We describe our techniques for generating traces and the actual design of our simulator in greater detail in \S\ref{sec:simulator}.

We ran our simulator on four image recognition models (described in greater detail in \S\ref{sec:model_char}), and we present evaluation results from these runs later in the paper. At a high level we found that in the typical case using fabric-based mechanisms to speed up training in the parameter server model has lower benefits than using host-based mechanisms that abandon the parameter server model in favor of other reduce strategies. We found that this held even when we combined both fabric-based mechanisms. Our basic conclusion is that optimizing communication for CNN training does not necessitate changes to the network fabric.

\section{Background}

In this section, we begin by discussing the computational model for CNN training. Following this, we provide an overview of communication paradigms for distributed CNN training. Overall, we provide overviews of the following mechanisms that we explore in this paper: in-network aggregation, IP multicast, ring-reduce, ring-reduce with multicast, and butterfly mixing. Next, we discuss how changes to the network fabric could be used in conjunction with the aforementioned communication paradigms -- and thus further accelerate training. 

\subsection{Distributed Training Steps}
\label{sec:computational:structure}

We now consider the computational process: how the model computation (forward pass and back propagation) interleave with the communication. We consider data parallel approaches to distributed learning, where each worker operates on the entire model but uses different training data. Data parallel learning is the most prevalent approach today. 

\subsection{Steps}
\begin{figure}
    \centering
    \includegraphics[width=0.4\textwidth]{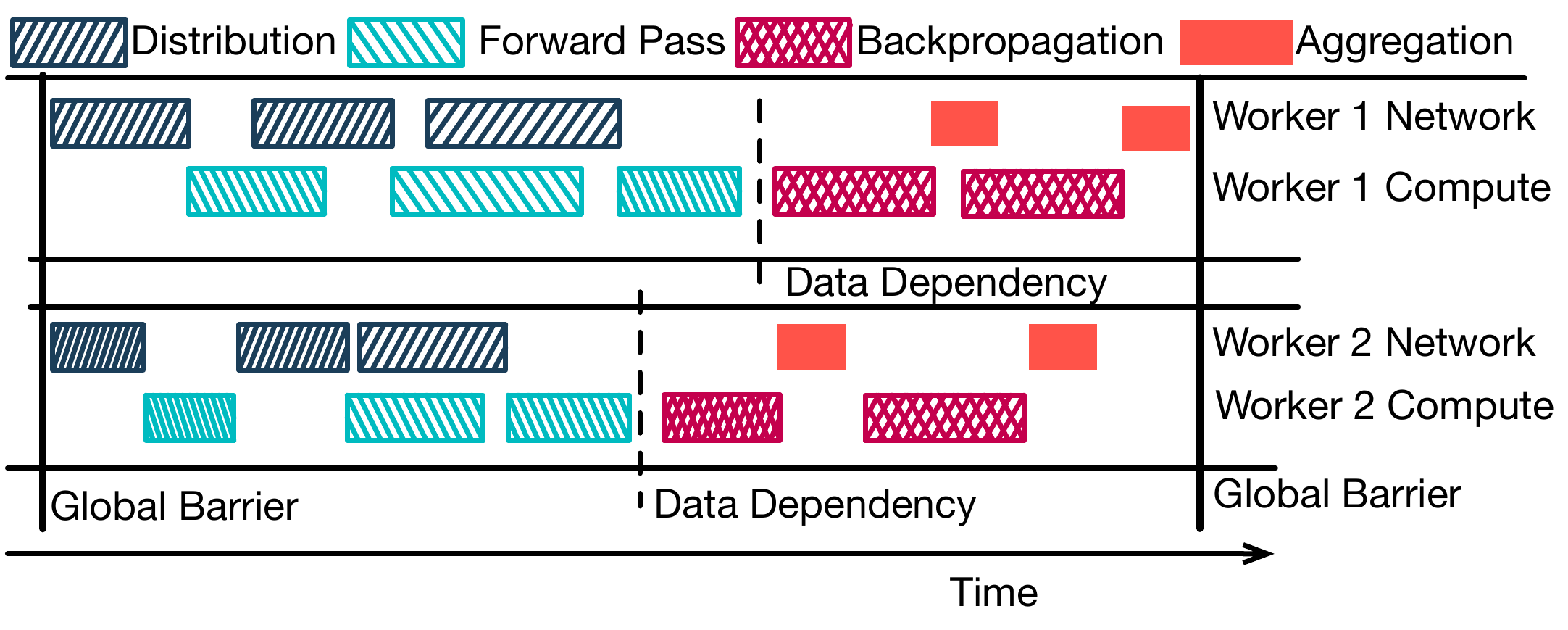}
    \caption{Steps in a distributed training job with param. server}
    \vspace{-0.2in}
    \label{fig:timeline}
\end{figure}

% There are four fundamental components of distributed CNN training.
Distributed CNN training proceeds in four steps:

\textbf{Distribution:} When using parameter servers, the parameter server updates to the worker constitutes the distribution phase. For butterfly mixing or ring-reduce, this would be the last communication phase, right before all workers receive the same updated model.

\textbf{Forward Pass:} Each worker selects a sample of the training data and uses the parameters to compute labels for this training sample. This computation is commonly referred to as the \emph{forward pass} since each layer operates on the training image in order. The training data used in this step is loaded concurrently during the distribution step and is not on the critical path. As a result, our analysis does not consider time taken loading training data.

\textbf{Backpropagation:} Next, each worker uses the supplied labels (from the training set) and computed labels to determine each layer's contribution to training error, and then computes an appropriate change to the layer. This computation is commonly referred to as \emph{backpropagation} and proceeds from the last layer in the neural net to the first one. Backpropagation utilizes results computed during the forward pass and, as a result of this data dependency, it cannot progress until the worker has finished the forward pass.

\textbf{Aggregation:} As the backpropagation progresses, the worker will send updates to a reducer. In the parameter server model, parameter servers are responsible for both aggregating and applying updates. An iteration is considered to have completed only when the parameter servers have received updates from \emph{all} workers, thus enforcing a global barrier (\ie across all workers) between the distribution and aggregation step. 

Within the parameter server model, each iteration of the algorithm consists of the four steps (shown in Figure~\ref{fig:timeline}). These four steps can be partially pipelined, in the following two ways. First, the forward pass proceeds layer-by-layer, and as a result a worker can begin the forward pass step as soon as it has received the parameters for the first layer of the CNN. Second, backpropagation also proceeds layer-by-layer, and workers can begin sending updates as soon as they have computed updates for a layer. To the best of our knowledge, all commonly used CNN frameworks employ pipelining to improve training performance. 

For end-host mechanisms (i.e., butterfly mixing and ring-reduce), the forward pass is not pipelined with the distribution phase. The back-propagation pipelining operates the same way.

\subsection{Algorithms for Efficient Reduction}
\subsubsection{Training on Parameter Servers}
\label{sec:background:ps_training}

One way to facilitate distributed CNN training makes use of one or more centralized parameter servers~\cite{Li2014ScalingDM}. These algorithms are implemented as a part of several CNN frameworks including TensorFlow~\cite{Dillon2017TensorFlowD}, Caffe2~\cite{Goyal2017AccurateLM}, and MXNet~\cite{Chen2015MXNetAF}. 

For parameter server based training, there can be two distinct, communication phases during each training iteration: (i) a distribution phase where CNN parameters are distributed to workers, which then execute a local training algorithm using these parameters, and (ii) an aggregation phase where each worker sends the local training algorithm's updates to one or more parameter servers.

The distributed CNN training algorithms we consider are iterative. Distributed training algorithms can be further classified into synchronous and asynchronous algorithms. \emph{Synchronous} training algorithms require that all workers agree on the model at the beginning of a training iteration; this is implemented by having the parameter server impose a barrier across workers. \emph{Asynchronous} training algorithms do not impose consistency requirements across workers. While asynchronous training algorithms decrease the time taken by each iteration, they increase the total number of iterations required and can thus slow down overall training time~\cite{Chen2016RevisitingDS}. Some companies like Google tend to favor synchronous training algorithms~\cite{Abadi2016TensorFlowAS}. In this paper we focus on synchronous training algorithms because the presence of synchronization barriers allows us to more easily reason about iteration time.

\subsubsection{Training without Parameter Servers}
\label{all_reduce_background}

Parameter servers present several issues to practitioners, which begin with selecting the correct ratio between the number of parameter servers and workers. 
Rather than aggregating parameter updates in centralized nodes, others have advocated efficient all-reduce algorithms that exchange parameters merely between worker nodes. In the context of training CNNs, two algorithms are most commonly discussed: ring-reduce~\cite{horovod} and butterfly mixing~\cite{DBLP:conf/sdm/CannyZ13}.

Ring-reduce, popularized by Uber's implementation (Horovod \cite{horovod}), requires that the workers connect in a ring. There are two communication phases. First, parameters in the model are assigned to each worker in a round robin fashion. In the first phase (analogous the aggregation phase), each worker begins computing gradient updates; when they complete the computation for the parameter assigned to it, it immediately sends the parameter to the next node in the ring. Upon receiving the update, each worker incrementally averages the update with it's locally calculated gradient. Once the last worker in the ring has received the parameter, it has the complete averaged model parameter from the entire cluster. In the second phase, this exact updated model is passed around the ring a second time such that all workers now possess this updated model. 

On the other hand, butterfly mixing performance scales logarithmically with the number of workers. At each phase within butterfly mixing, each worker simultaneously sends the entirety of its model to one other worker. After receiving an update, the worker averages its local model with the received model. Suppose there are four workers: $A$, $B$, $C$, $D$. In the first phase of communication, $A$ and $B$ simultaneously exchange their entire model, while $C$ and $D$ follow suit. At this point, $A$ and $B$ contain the same averaged model, as does $C$ and $D$. In the second phase of communication, $A$ sends its averaged model to $C$ (and vice versa), while $B$ sends its averaged model to $D$. This concludes the all-reduce. Butterfly mixing reduces the number of communication phases required during updates at the cost of sending a larger amount of data over the network fabric.

\subsection{Mechanisms to Accelerate Training}
\label{subsection:mechanisms}
The network fabric is capable of providing support for distributed training at multiple levels. Here, we discuss two proposals and how they can used in conjunction with the all reduce algorithms presented earlier.

\subsubsection{IP Multicast}
Multicast ~\cite{RFC1112} implementations are designed to be bandwidth efficient. For CNN training, multicast ensures that the amount of data traversing any network link does not scale with worker count. In practice, deploying multicast requires addressing a litany of considerations (e.g., membership, discovery, reliability). However, we do not address them in this paper, noting only that CNN training tasks involve bulk transfers (making reliability easier to solve since one has time to identify and recover from errors, and there are various reliable multicast implementations available) and tend to run over several hours with a constant set of workers, simplifying the management problems.

When using a parameter server, IP multicast assists solely with the distribution phase. During the distribution phase each parameter server sends each worker all of its parameters -- $w$ copies of the same data when deployed in a cluster with $w$ workers. Enabling IP multicast would allow each parameter server to send a single copy of its parameters. It is also possible to use IP multicast with ring-reduce during the second ring during it's model distribution.

\subsubsection{In-network Aggregation}
In-network aggregations improve training time when using parameter servers. The data sent during the aggregation phase varies by worker. To reduce traffic in this phase, the network needs to implement the model's aggregation semantics. This is possible either through the use of software switches operating on an overlay network~\cite{Mai2015OptimizingNP} or through the use of programmable switch ASICs~\cite{Sapio2017InNetworkCI} such as Barefoot Tofino~\cite{barefoot}. In this approach the network buffers worker updates and aggregates them before sending them to parameter servers. The benefits of this approach mirror those achieved through multicast; this mechanism also presents several deployment challenges including requiring the use of new, specialized switching hardware, allocation of compute and memory resources on switches, and mechanisms for isolating CNN training traffic. We also do not address these issues in the paper, and refer the interested reader to recent discussions on this topic~\cite{Sapio2017InNetworkCI}.

\section{Analytical Models Insufficient}
\label{sec:analytical_insufficient}

Our initial attempts to derive an analytical model for training performance provided \textit{intuition} about mechanism scaling, but failed to identify the reasons about the relative performance of the various mechanisms. In particular, it fails to capture two specific components of CNN training that significantly affect performance.

\textbf{First, the backpropagation phase consists of fine-grained, causal, interleavings between compute and network}. This set of interleavings is highly \textit{model specific}. Each model has \textit{hundreds} of composite operations that are interspersed over highly uneven units of time. Moreover, radical new CNN proposals appear frequently; thus developing a novel analytical model for each new CNN design is highly burdensome. Rather than attempt to model this complex interaction, our simulator uses a comprehensive set of empirical traces that are easy to collect.

\textbf{Secondly, different phases of computation (e.g., distribution and aggregation) overlap non-deterministically.} Recall that while there is a \textit{global barrier} at the parameter server, each worker unit has a \textit{local barrier} before it initiates backpropagation. If we assume 1.) every local barrier is reached simultaneously and 2.) there is no variance in the compute part of backpropagation, this would -- within the constraints of our computational model -- maximize the amount of incast on the parameter server. On the other hand, any delta between workers hitting the local barrier would reduce each worker's overlap between the backpropagation phases; this delay reduces incast. We refer to this delta as \textbf{backpropagation staggering}. Backpropagation staggering is influenced by two factors that are extremely difficult to model analytically. First, it requires reasoning across both the overlapping distribution and aggregation phases. The mechanism used to distribute parameters, as well as the qualities of the individual parameters themselves (i.e., how long it takes to compute and send over the network) affect the amount of backpropagation staggering. Second, there is natural variation in worker processing time that further influences the staggering. This is not limited to the parameter server framework; in the case of end-host mechanisms like butterfly mixing, pipelined parameter mixing between workers can interfere with each other. In Section \ref{sec:eval}, we will show in much greater detail how the amount of backpropagation staggering influences how the mechanism and model performs. Thus, an analytic model which fails to capture this phenomenon will not be sufficient to answer the questions we posed.

\section{Trace-driven Simulator Design}
\label{sec:simulator}

We seek to develop a simulator that captures the performance nuances described in Section \ref{sec:analytical_insufficient}, while simultaneously being sufficiently flexible to accommodate a limited scope of potential changes in feedforward CNN models and Tensorflow. Thus, we moved towards developing a trace driven simulator which we describe in greater detail in this section.

To ensure that the trace collection is simple and effective, a trace must have two attributes. First, it must be network agnostic, so the same trace can be used for different network settings. Second, it must accurately model pipelining. The traces were generated by adding minor instrumentation to TensorFlow 1.4. Computation in TensorFlow is represented as a dataflow graph, where individual \textit{operations} are represented as nodes on the graph, while parameters (i.e., tensor) are transferred along the edges of the graph. Our instrumentation records \textit{send} operations along the dataflow graphs. The trace is very simple. For each operation, it shows five attributes. 1.) event time 2.) parameter name -- identified by an edge name (e.g., \textit{conv1/weights/read} tells us that the first convolution operation is being read by the worker) 3.) size of the parameter ready to be queued on to the network. 4.) source device (e.g., worker0) 5.) destination device (e.g., parameter server). This trace is simple to generate, parse, and possibly even modify if the operator wants to run simulations on synthetic models.

For each neural net model, we collected a different representative trace. On each iteration, we partitioned the events into two traces: an aggregation trace and a distribution trace. 

For the aggregation trace, we identified specific send operations that are \textit{gradient} operations triggered automatically by TensorFlow's SyncReplicaOptimizer. The aggregation trace allows us to accurately model how the parameters are sent over the network. Recall from the computational model that due to extensive pipelining, the calculated gradients are delivered to the parameter server as soon as the parameter gradient is calculated. In the gathering of this data, note that not every operation from the worker to the parameter server falls on the critical path. The beginning of the aggregation is marked by a \textit{dependency operation} that varies from model to model. We only consider the send operations that occur following the dependency operation, since all other send messages prior to that dependency operation are pipelined with the worker's forward pass; thus, they do not fall on the critical path. For example, it is common for CNN training to use \textit{batch normalization}, which normalizes the inputs to each layer. The worker will send the result of its batch normalization computation to the parameter server so the PS can compute moving averages over multiple iterations. Such operations must be removed from the aggregation trace. There are other operations that must be filtered out that arise occasionally on a per-model basis (such as the use of \textit{auxiliary logits}).

For the distribution trace, we used the TensorFlow timeline tool to identify the particular send operations originating from the parameter server that triggered forward pass operations on the worker. The purpose of the distribution trace is to obtain the \textit{order} in which parameters are queued on the network. We demonstrate in the evaluation (\S\ref{sec:eval}) how the ordering of parameters affects the end-to-end performance of distributed training. Additionally, we profiled the forward pass time on a single GPU. We later use this for emulating the pipeline effects in the forward pass, which we find plays an insignificant role in typical use cases; this is due to the one-to-many communication overhead arising from the parameter server.

To simulate normal performance computational performance variance across GPUs, we recorded traces for each worker in clusters of difference sizes and across several distinct clusters. To ensure that the trace is agnostic to the network and size of the cluster, the recorded time of a trace event is relative to the first event in that trace. Different workers in a cluster finish receiving the model at different times due to network topology and parameter ordering. The absolute times in the aggregation trace are affected by the network conditions, so aggregation times are recorded relative to the first aggregation event. Thus, we are able to simulate the network effects across a wide variety of cluster sizes.

\begin{table}[t]
\centering
\footnotesize
% \resizebox{\columnwidth}{!}{
\begin{tabular}{@{}lrrrr@{}}
\toprule
\multicolumn{1}{c}{\textbf{CNN Name}} & \multicolumn{1}{c}{\textbf{1 PS}} & \multicolumn{1}{c}{\textbf{2 PS}} & \multicolumn{1}{c}{\textbf{4 PS}} & \multicolumn{1}{c}{\textbf{8 PS}}\\ \midrule
VGG-16 Sim             & 21.0          & 22.5  & 19.3       & 18.2  \\
VGG-16 Real            & 22.5          & 22.8  & 20.8 & 19.3  \\
Inception-v3 Sim       & 2.29        & 2.29  &  1.37 & 0.852 \\
Inception-v3 Real      & 2.16          & 2.16  & 1.49 & 1.3 \\
Resnet-200 Sim         & 7.15        & 3.34  & 2.3  & 2.29           \\
Resnet-200 Real        & 5.89        & 2.3  & 1.71  & 1.71           \\
Resnet-101 Sim       & 4.57       & 2.37     &  1.52 & 1.5 \\ 
Resnet-101 Real       & 3.7      & 1.58     &  0.855 & 0.9 \\ 
\end{tabular}%}
\caption{Comparison of measured (Real) iteration time compared to simulation prediction times (sim) on a cluster of 8 workers}
\label{tab:validation_table}
\vspace{-0.2in}
\end{table}

\section{Simulator Validation}
We validated our simulator by comparing the results of the simulator against actual runs. In a cluster of 8 workers, our results are shown in Table~ \ref{tab:validation_table}. For the most part, our simulation results accurately predict the performance trend with more parameter servers. In many of these cases that use multiple parameter servers, the CNN weights are not evenly distributed among the parameter servers, causing the performance improvements to plateau. Our simulation effectively reflects this behavior. There are two notable points at the far ends of the spectrum -- Inception-v3 with 8 PSs, and Resnet-200 with 1 PS -- where our simulation fails to match our empirical measurement, although both still capture the general scaling trend. There are several possible explanations for this. First, our simulation makes an assumption that parameters in the distribution phase occurs in a round-robin fashion over the workers. While this assumption is sufficient for most of the settings that we looked at, our observations of the actual distribution send-traces reveal that there are some minor overlaps in the way that worker parameters are being sent. This suggests that the round robin assumption is slightly stronger than reality -- especially in the case of 1 PS where the amount of backpropagation staggering is most pronounced.

\section{Model Characterization}
\label{sec:model_char}

\begin{table}[t]
    \centering
    \footnotesize
 \resizebox{\columnwidth}{!}{
    \begin{tabular}{@{}lrrrr@{}}
    \toprule
    \multicolumn{1}{c}{\textbf{CNN Name}} & \multicolumn{1}{c}{\textbf{\# Layers}} & \multicolumn{1}{c}{\textbf{\# Weights}} & \multicolumn{1}{c}{\textbf{Model Size (Gb)}} &
    \multicolumn{1}{c}{\textbf{\# of FLOPs}}\\ \midrule
        Inception-v3 & 21 & $2.5 \times 10^6$ & 0.715 & $1.1\times 10^{10}$ \\
        VGG-16 & 22 & $1.9\times 10^8$ & 6.58 & $5.4 \times 10^{10}$ \\
        Resnet-101 & 103 & $2.2 \times 10^6$ & 1.42 & $1.3 \times 10^{10}$ \\
        Resnet-200 & 202 & $2.2 \times 10^6$ & 2.06 & $2.8\times 10^{10}$ \\    
    \end{tabular}}
    \caption{Complexity of the CNN models we considered, including both weight and pooling layers in our layer count.}
    \vspace{-0.1in}
    \label{tab:model_basic_attributes}
\end{table}

\begin{table}[t]
    \centering
    \footnotesize
 \resizebox{\columnwidth}{!}{
    \begin{tabular}{@{}lrrrr@{}}
    \toprule
    \multicolumn{1}{c}{\textbf{CNN}} & 
    \multicolumn{1}{c}{\textbf{Fwd Pass Comp}} & 
    \multicolumn{1}{c}{\textbf{Bkprop Comp}} & 
    \multicolumn{1}{c}{\textbf{Bkprop Net, 25~Gbps}}& 
    \multicolumn{1}{c}{\textbf{Comp:Net Ratio}}\\
    \midrule
        Inception-v3 & 0.176 sec & 0.296 sec & 0.028 sec & 10.6 \\
        VGG-16 & 0.169 sec & 0.024 sec & 0.263 sec & 0.09\\
        Resnet-101 & 0.176 sec & 0.180 sec & 0.052 sec & 3.46\\
        Resnet-200 & 0.357 sec & 0.34 sec & 0.082 sec & 4.14
    \end{tabular}
}
    \caption{Compute and network times during the backpropagation of the model. Note that the backprop compute time does \textit{not} include the time to calculate the first layer of backpropagation.}
    \vspace{-0.1in}
    \label{tab:model_backprop_interspersion}
\end{table}

Which \textit{CNN model characteristics} influence a mechanism's performance benefit? We've identified four such model attributes that impact the extent to which acceleration mechanisms will benefit performance and show where our CNN models fall on those dimensions. Other, non-deterministic factors not related to the CNN characteristics may influence training performances, but are discussed earlier in Section \ref{sec:analytical_insufficient}. With the exception of the forward pass, these attributes are all extracted directly from the trace. We discuss this in the context of the four distinct, commonly deployed image classification models we deployed and analyzed: Inception-v3, Resnet-200, Resnet-101, and VGG16. They were trained using ImageNet data.  \\
\noindent\textbf{Distribution of parameter sizes over model, in particular the size of the last parameter}. Many models exhibit a very parameter heavy fully connected last layer, which represents a significant fraction of the model size. Inception-v3 and VGG16 both possess very memory expensive fully connected layers, while Resnet-200 and Resnet-101 are relatively even throughout. \\
\noindent\textbf{Computation/network bottleneck \textit{after} the first layer of back-propagation.} For all future references to backpropagation compute/network ratio, the first layer of backpropagation compute is not included. Once the first layer of backpropagation has been calculated, how interspersed are the computational and network elements of a CNN model? Table \ref{tab:model_backprop_interspersion} shows the amount of time spent in communication and computation, and a \textit{compute:net} ratio. VGG16 spends nearly all it's computational time calculating the first back propagation parameter, thus exhibiting the smallest compute:network ratio. On the other hand, Inception-v3, a model which is similarly skewed, is compute intensive even after the first layer of back propagation is computed. \\
\noindent\textbf{Forward Pass Time}: See Table \ref{tab:model_backprop_interspersion}\\
\noindent\textbf{Raw size of the model}. We show model sizes in Table \ref{tab:model_basic_attributes}. They range from very large (6.58Gb) to small (0.7Gb)\\
As we will show in Section \ref{sec:eval}, the first two characteristics listed above heavily influence the amount of backpropagation staggering, as they both affect the overlapping distribution and aggregation phases.

\vspace{-.1in} 
\section{Evaluation}
\label{sec:eval}

In this section, we evaluate the efficacy of the mechanisms described in previous sections, and explore how the CNN model characteristics interact with the mechanism. Finally, we explain how they jointly impact performance. We also present head to head comparisons of competing mechanisms. Finally, we show that our rankings and intuitions generalize to two types of future training conditions: 1. larger models and 2. faster processors.

The traces used in the simulations were derived from clusters in AWS EC2 running Tensorflow 1.4, with a fixed batch size of 32 training instances per worker unit. 

Our simulations show that in all cases, an end-host mechanism (ring-reduce) offers performance improvements greater than or equal to any in-network mechanism. Thus, rather than having to make a trade-off between performance and infrastructure cost, operators can instead deploy software mechanisms running on the end-host and expect to get equal or better performance. 

\subsection{In-Network Optimization}
\label{sec:mcast_vs_agg}

We look at network fabric optimizations  -- in-network aggregation and multicast -- which primarily are used to accelerate training \textit{when using a parameter server.}. 

\begin{figure}
    \centering
    \includegraphics[width=0.44\textwidth]{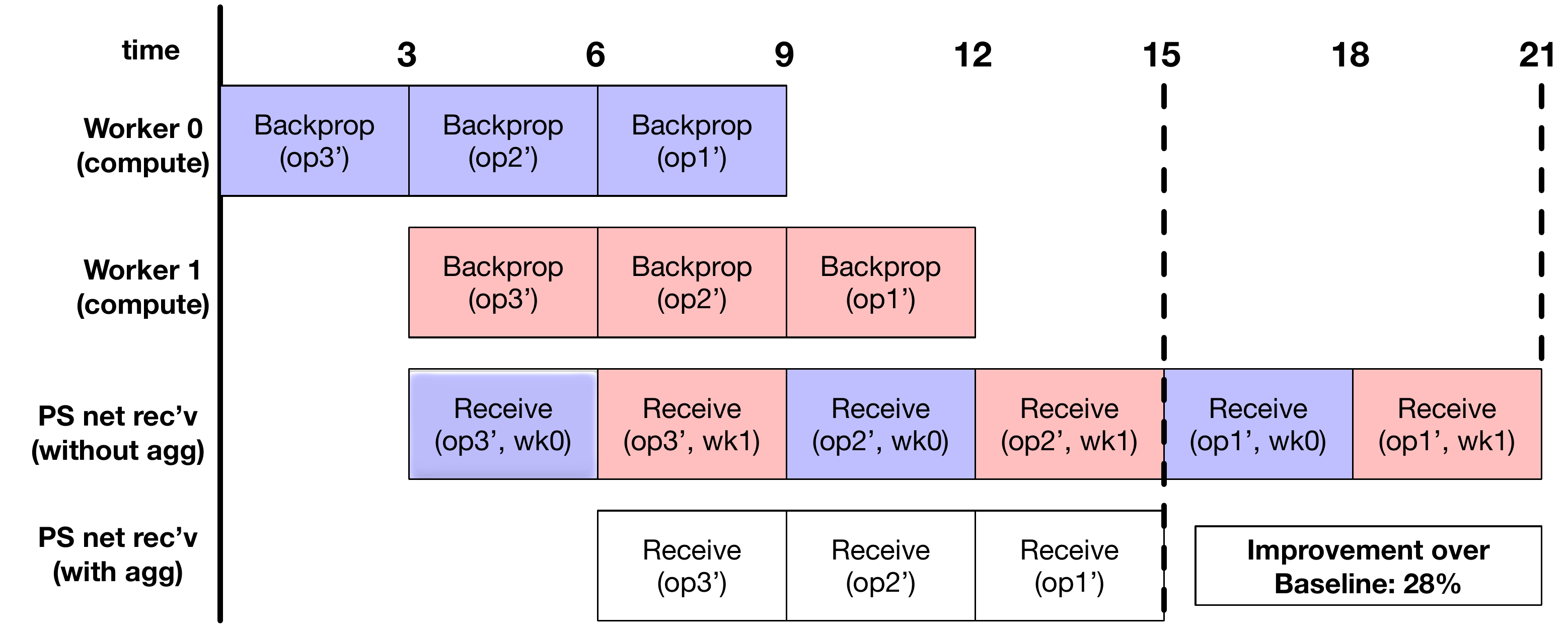}
    \caption{Aggregation phase of \textbf{computationally/communication even}, 3 layer (op) toy model, but with staggered backpropagation start times. PS = parameter server}
     \label{fig:delay_comparison_agg}
     \vspace{-.1in}
\end{figure}

\begin{table}[t]
    \centering
    \footnotesize
 \resizebox{\columnwidth}{!}{
    \begin{tabular}{@{}lrrr@{}}
    \toprule
    \multicolumn{1}{c}{\textbf{Model Name}} & \multicolumn{1}{c}{\textbf{Aggregation Only}} & \multicolumn{1}{c}{\textbf{Multicast Only}} &
    \multicolumn{1}{c}{\textbf{Multicast + Aggregation}}\\ \midrule
        Inception-v3 & 1.34x & 1.69x & 3.28x \\
        VGG-16 & 1.89x & 1.94x & 22.0x \\
        Resnet-101 & 1.65x & 1.79x & 6.07x \\
        Resnet-200 & 1.65x & 1.85x & 6.7x \\
    \end{tabular}}
    \caption{Factor speedup of network support models relative to a baseline model with no network support. 32 workers, 25~Gbps}
    \vspace{-0.1in}
    \label{tab:network_support_speedup}
\end{table}

\subsubsection{In-network Aggregation}~\\
\label{sec:agg_analysis}
\noindent\textit{Factor 1: Large Last layer of CNN \textit{reduces} impact}: In section \ref{sec:analytical_insufficient}, we defined backpropagation staggering. Recall that increased backpropagation staggering reduces incast, which consequently reduces the impact of in-network aggregation. The amount of backpropagation staggering is positively correlated with the time taken to execute the the penultimate layer(s) of the CNN. During the distribution phase, parameters are distributed between workers in a round robin manner. Thus, the forward pass on each worker is blocked until that single parameter (which can be over 5~Gb in the case of VGG16) reaches the worker in its entirety. Larger final parameter(s) causes each worker node to begin it's individual backpropagation at increasingly staggered times. To illustrate this, we consider a simple example with a 3 operation CNN model. Each operation takes three seconds to compute and three seconds to send over the network. In the case where all the workers start backpropagation simultaneously (not shown), aggregating the new model takes 21 seconds. When using in-network aggregation, aggregation takes 12 seconds (a 43\% improvement). In contrast, Figure~\ref{fig:delay_comparison_agg} shows aggregation in the same setup when backpropagation start-time is staggered between workers. While performance \textit{without} in-network aggregation stays constant (21 seconds), in-network aggregation only improves performance by 28\%. Why? In-network agg reduces parameter server network bottleneck by aggregating parameter updates from \textit{all workers}. This aggregation can't proceed until the \textit{last worker} has communicated that parameter update. 

Both VGG16 and Inception-v3 have large last layers that will further stagger the time at which each worker begins backpropagation, making them particularly susceptible to this effect.

\noindent\textit{Factor 2: Network dominated backpropagation time \textbf{increases} impact}: Recall that during the back-propagation process, parameters updates are sent as soon as they are calculated. If the calculation time between parameter updates is relatively short, there will be fewer parameter updates queued inside the network to be sent to the parameter server.  Assessing the degree to which the backpropagation is network-bound cannot be evaluated solely by the overall time of backpropagation. VGG16 has the longest \textit{overall} backpropagation time, yet enjoys the largest percentage improvement in performance from in-network aggregation. For VGG16, the majority of the backpropagation computation is spent on computing the \textit{first backpropagation layer}, an expensive fully connected layer. Once this first computational step completes, the remaining model parameters are quickly calculated and queued in the network. Conversely, Inception-v3 spends a significant amount of time doing backpropagation computation even after the first backprop layer is computed. 

\textit{Results:} For 32 workers and 25~Gbps network, Table \ref{tab:network_support_speedup} demonstrates the performance improvements derived from using just in-network aggregation to accelerate distributed training. Inception-v3 (which has a compute-intensive backpropagation) experiences the least performance gain from in-network aggregation, while VGG16 (which has a network-intensive backpropagation) experiences the most.

\vspace{-.1in}
\subsubsection{Multicast}~\\
\label{sec:mcast_analysis}
\noindent\textit{Factor 1 -- Model Size increases impact} Unlike in-network aggregation, the distribution phase is initiated with a global barrier. All the parameters are ready simultaneously, so there is no pipelining on the send-side as in the case of in-network aggregation. Thus performance gains from multicast are more directly a function of model size. 

\noindent\textit{Factor 2 -- Forward Pass Pipelining decreases impact (slightly)} As the worker receives model parameters, it partially executes the forward pass. However, the forward pass is unlikely to be the bottleneck in the distribution phase, especially as the number of workers grows. Recall that the simulated parameter server distributes parameters to the workers in a round robin fashion. More workers will allow for more computation time between parameter distributions. Consequently, forward pass pipelining has a very slight (if any) impact on multicast performance. 

\noindent\textit{\textit{Non-Factor} -- decreased backpropagation staggering}: When using multicast, workers receivers parameters at roughly the same time (within minimal link latency). Thus backpropagation staggering is decreased, as we expect workers to initiate backpropagation simultaneously (approximately). Does the resultant increased incast hurt iteration performance? We find this not to be the case unless the following condition is met. Let $D$ be the delay between worker backpropagation start times, $B$ be the \textit{full backpropagation time} (i.e., including both the computation and network transfer time), and $C$ be the compute time for the backpropagation time of the first layer. In order for the decreased backpropagation staggering to give back performance, the following must hold: $D > B - C$. This is highly unlikely in the multicast case; multicast should beget a very minimal $D$.

\textit{Results}: Again, we refer to Table \ref{tab:network_support_speedup}. Multicast impact is more directly proportional to the size of the model. Resnet-101 (1.78x) and Resnet-200 (1.85x) are both larger than Inception-v3 and smaller than VGG16. Their performance gains likewise fall between the performance gains of those models.

\subsubsection{Head to Head: In-network Aggregation vs. Multicast}
\label{sec:mcast_agg_h2h}
As shown in Table \ref{tab:network_support_speedup}, multicast outperforms or approximates the performance gains from using in-network aggregation in all cases. Later, we will show that both approaches are weaker than end-host based acceleration mechanisms, but we briefly provide intuitions for why multicast is more effective than in-network aggregation. Fundamentally, in-network aggregation is tied to backpropagation and multicast is tied to the forward pass. Generally speaking, the forward pass is significantly faster than the backpropagation; refer to Table \ref{tab:fwd_pass_benchmarks}. Moreover, the extent to which the forward pass is bottlenecked on compute decreases with more workers. On the other hand, the compute:network ratio of backpropagation remains constant with the amount of workers. In fact, increasing workers only leads to staggered backpropagation start times, which reduces the benefits of in-network aggregation. \textbf{Multicast individually is more impactful than in-network aggregation alone.} Table \ref{tab:network_support_speedup} indicates that multicast provides larger performance gains than in-network aggregation across the board.

\begin{table}[t]
\centering
\footnotesize
% \resizebox{\columnwidth}{!}{
\begin{tabular}{@{}lrrr@{}}
\toprule
\multicolumn{1}{c}{\textbf{CNN Name}} & \multicolumn{1}{c}{\textbf{GPU Model}} & \multicolumn{1}{c}{\textbf{Forward Pass}} & \multicolumn{1}{c}{\textbf{Backprop}}\\ \midrule
Resnet-200            & Maxwell Titan X          & 170 ms & 384  ms  \\
Resnet-200            & Pascal Titan X          & 315 ms & 520 ms\\
VGG-16            & Maxwell Titan X          & 173 ms & 416 ms  \\
VGG-16            & Pascal Titan X          & 98.2 ms & 260  ms \\
Resnet-101       & Maxwell Titan X        & 109 ms & 190 ms \\
Resnet-101      & Pascal Titan X          & 162 ms & 258 ms \\
Inception-v1       & Maxwell Titan X        & 91.3 ms & 141 ms \\
Inception-v1      & Pascal Titan X          & 57.5 ms & 85.9 ms \\
\end{tabular}%}
\caption{Forward Pass vs. Backpropagation Benchmarks~\cite{cnnbenchmark}}
\label{tab:fwd_pass_benchmarks}
\vspace{-0.2in}
\end{table}
        
\subsubsection{Multicast \textit{plus} in-network aggregation}~\\
\label{sec:mcast_with_agg}
Finally, what happens when multicast is \textit{combined} with in-network aggregation? Because multicast supports the distribution phase and in-network aggregation supports the aggregation phase, both approaches can be simultaneously used to improve performance. In fact, multicast \textit{puts in-network aggregation in the best position to succeed} because it strongly decreases backpropagation staggering. Table \ref{tab:network_support_speedup} shows clearly that using multicast with in-network aggregation yields substantially better performance gains than using either just multicast or just in-network aggregation.

\textit{Results:} Multicast plus in-network aggregation benefits VGG16 the most, as the performance gains increase from ~1.9x to 21.2x. For all models, using both multicast with in-network aggregation results in \textit{more than additive} performance gains from the individual mechanisms, for reasons described in the previous paragraph.

\subsubsection{Summary of In-network Changes}
\label{sec:ps_eval_summary}
Based on just looking at in-network mechanisms, we find that optimizations can be ranked as: multicast + aggregation, multicast, aggregation. This leads us to conclude that if one is required to use the parameter server model, then using multicast jointly with in-network aggregation yields the largest performance improvements. While both multicast and in-network aggregation offer their own set of deployment challenges, multicast outperforms in-network aggregation in the CNNs we tested. As evidenced by the CNN characteristics that factored into each mechanism's impact, the interaction between the distribution phase and aggregation phase \textit{across all workers} is key. While the aggregation phase holds more complexity in terms of network/compute interleavings, it is actually the distribution phase optimizations which dictate how much the aggregation phase sits on the critical path. Future work should not solely evaluate the efficacy of accelerating either the distribution phase or aggregation phase in isolation.

\begin{table}[t]
    \centering
    \footnotesize
 \resizebox{\columnwidth}{!}{
    \begin{tabular}{@{}lrrr@{}}
    \toprule
    \multicolumn{1}{c}{\textbf{Model Name}} & \multicolumn{1}{c}{\textbf{Ring-Reduce}} & \multicolumn{1}{c}{\textbf{Ring-Reduce + Multicast}} &
    \multicolumn{1}{c}{\textbf{Butterfly Mixing}}\\ \midrule
        VGG-16 & 24.6x	& 24.6x	& 11.3x \\
        Resnet-200 & 6.75x	& 6.76x	& 6.79x \\
        Resnet-101 & 6.55x	& 6.71x	& 6.46x \\
        Inception-v3 & 3.35x	& 3.41x	& 3.41x \\
    \end{tabular}}
    \caption{Ring-reduce is most effective when parameters are evenly distributed, while butterfly mixing performs well at 25~Gbps.}
    \vspace{-0.2in}
    \label{tab:all-reduce-speedup}
\end{table}

\subsection{End-Host Mechanisms}
We analyze two all-reduce algorithms: Horovod ring-reduce and butterfly mixing. Table 6 shows speedup over baseline for both algorithms when run with 32 workers at 25~Gbps links. The table also shows improvements when ring-reduce is combined with multicast.

\subsubsection{Ring-reduce}~\\
\label{sec:ring_reduce}
Recall that in ring-reduce, the parameters are assigned to workers round robin. One critical issue that must be addressed when doing this is that often a significant fraction of a model’s raw size comes from a single parameter (e.g., VGG16, Inception-v3). Analytically, the communication overhead of ring-reduce is $2(W-1)*(max\ parameter)$ where $W$ is the worker count. This overhead can be prohibitively large when a model has a single huge parameter, e.g. VGG16's 5.4~Gb fully-connected layer. This is consistent with our simulation with 32 workers at 10~Gbps where ring-reduce iteration time was 34.0 seconds. CNN models (especially those ending in a fully connected layer) tend to have a few layers that take up a significant percentage of the overall model size; thus even an optimal assignment of model parameters would still result in parameter size imbalances for a worker on a ring.
    
To address this,  we modified our simulator to use parameter messaging, where parameters are partitioned evenly between workers (discussed more in \S\ref{sec:message_pipelining}). In the rest of this section, all ring-reduce results make use of this messaging mechanism. What model characteristics positively influence ring-reduce performance relative to baseline?

\textit{Factor: Network-dominated backpropagation increases impact}: For ring-reduce, each worker begins backpropagation at the same time. This is imposed by a global barrier. In the optimal case for ring-reduce, each worker sends it’s assigned model parameter updates at the same time. Then, every link in the ring would be nearly identically utilized and network contention would be minimized. However, each worker can only send its model parameter update when that gradient has been calculated. Longer time for backpropagation computation on each parameter results in deviation from the optimal case. Table \ref{tab:all-reduce-speedup} shows the model with the most compute-bound back-propagation, Inception-v3, has the lowest performance improvement from ring-reduce (3.3x). In contrast, VGG16, the model with the largest performance improvement (24.6x), has the most network-bound back-propagation process.

\subsubsection{Butterfly Mixing}~\\
\label{sec:butterfly_mixing}
Recall that the first phase of butterfly mixing merely involves a forward pass. At the point the forward pass begins, each worker already received the complete set of model parameters. Once each parameter is calculated, it will be sent $log(W)$ times, where $W$ is the number of workers. While this communication is sequential, this process can be pipelined between workers. What factors impact performance for butterfly mixing?

\textit{Factor: Compute dominated backpropagation increases impact:} During backpropagation, longer gradient computations between parameters gives workers a chance to pipeline communications among the $log(W)$ steps. However, models with network dominated backpropagation will \textit{still experience a substantial boost with butterfly mixing}; for example, Table \ref{tab:all-reduce-speedup} indicates that VGG16 still gets a 11.3x speedup. 

\begin{figure*}
\centering
\begin{minipage}{.3\textwidth}
    \includegraphics[width=\textwidth]{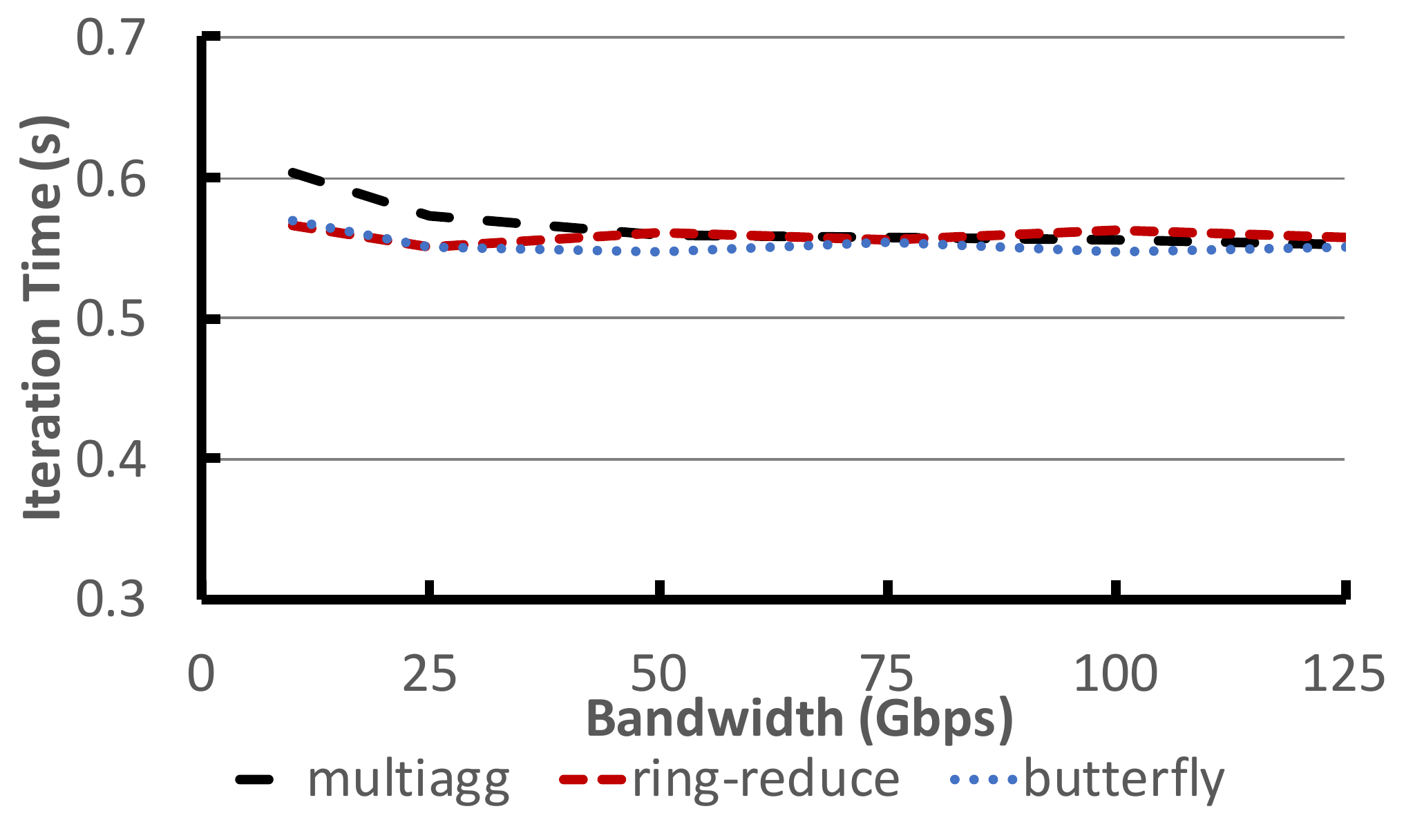}
    \caption{\textbf{Varying Bandwidth}: Mechanism Rankings for Inception-v3 training time on 32 workers}
    \label{fig:inception_overall_ranking}
\end{minipage}
\hfill
\begin{minipage}{.3\textwidth}
    \includegraphics[width=\textwidth]{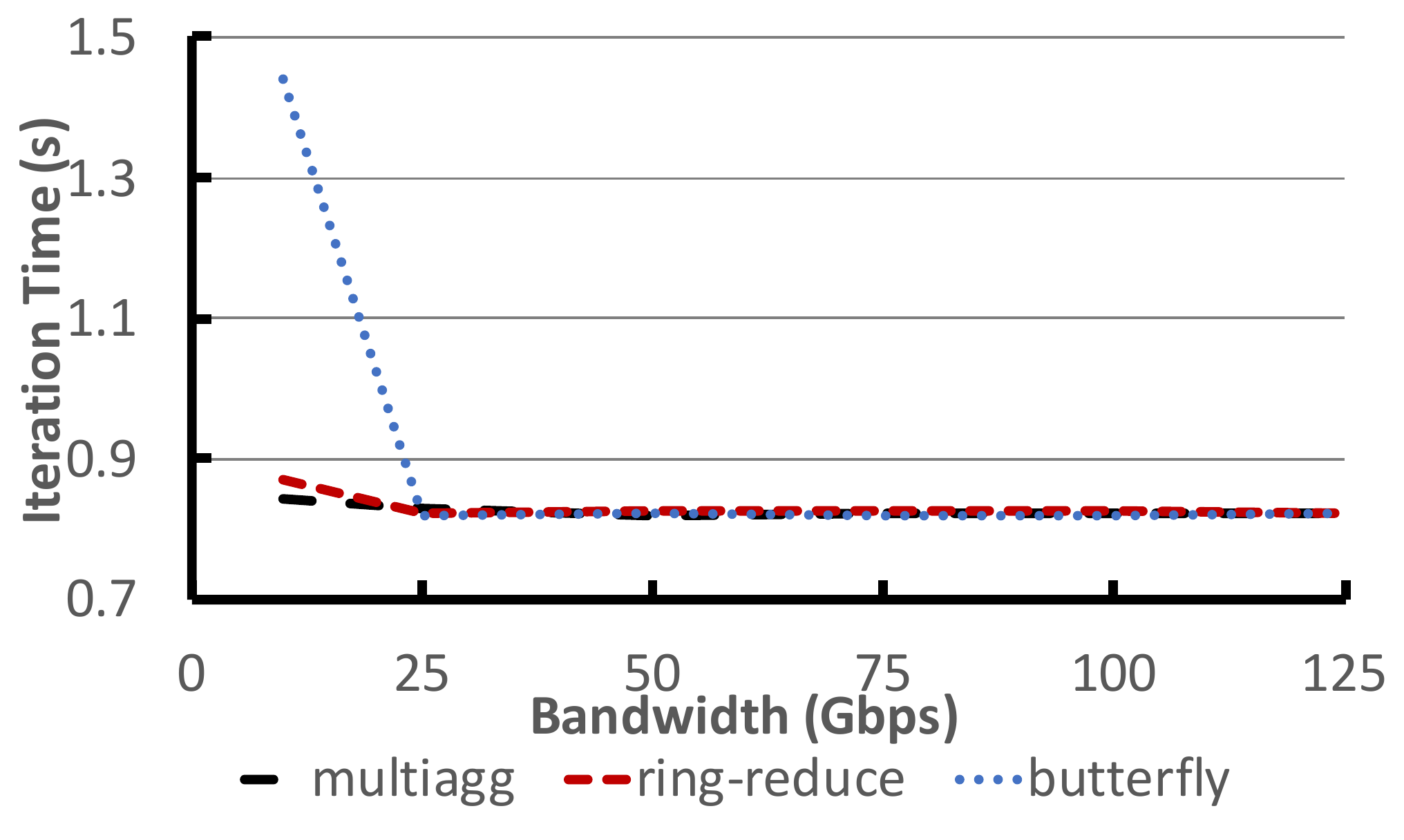}
    \caption{\textbf{Varying Bandwidth}: Mechanism Rankings for Resnet-200 training time on 32 workers}
    \label{fig:resnet200_overall_ranking}
\end{minipage}
\hfill
\begin{minipage}{.3\textwidth}
    \includegraphics[width=\textwidth]{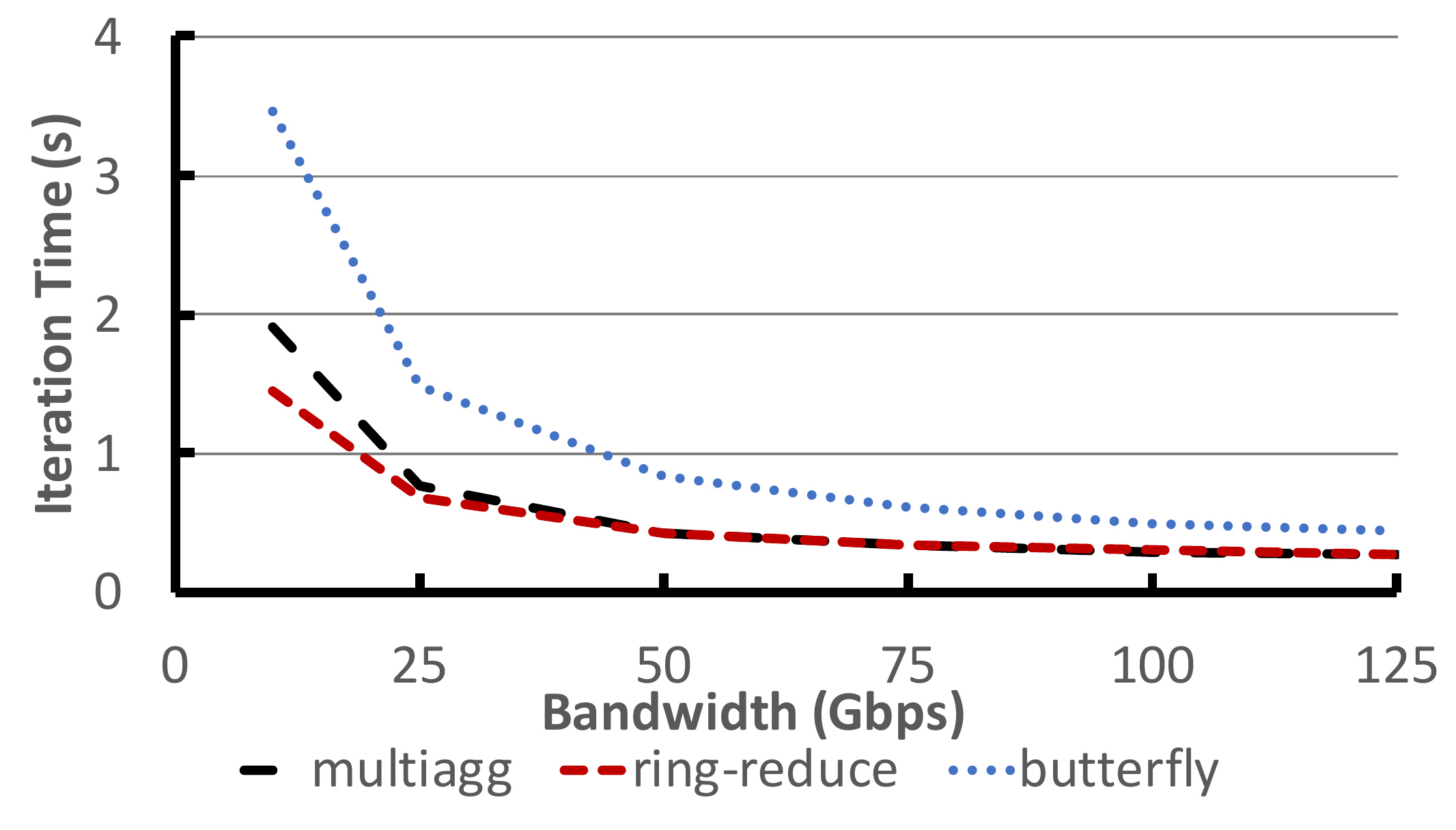}
    \caption{\textbf{Varying Bandwidth}: Mechanism Rankings for VGG16 training time on 32 workers}
    \label{fig:vgg_overall_ranking}
\end{minipage}
\end{figure*}

\begin{figure*}
\centering
\begin{minipage}{.3\textwidth}
    \includegraphics[width=\textwidth]{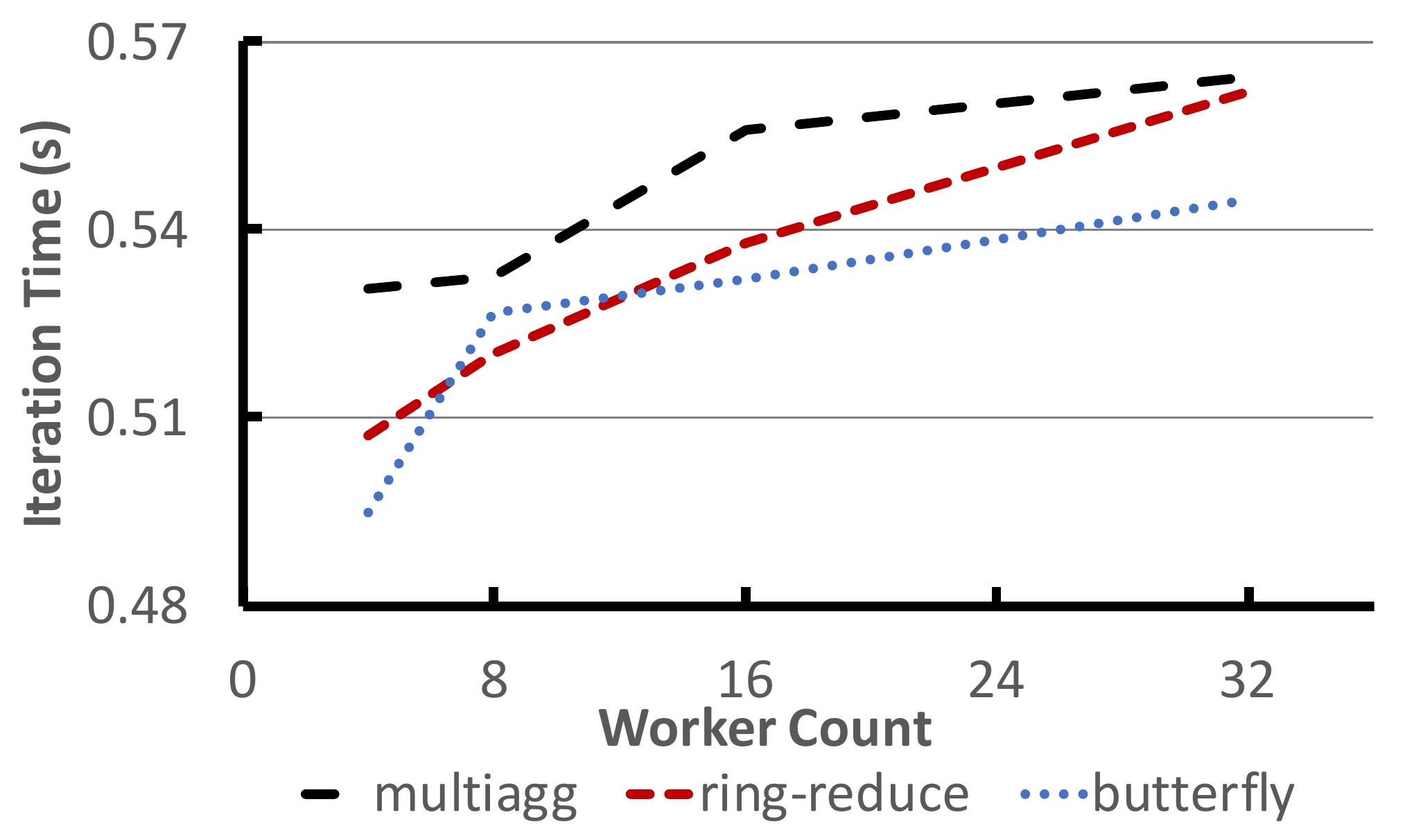}
    \caption{Varying Worker Count: Inception-v3 performance at 25~Gbps}
    \label{fig:inception_worker_ranking}
\end{minipage}
\hfill
\begin{minipage}{.3\textwidth}
    \includegraphics[width=\textwidth]{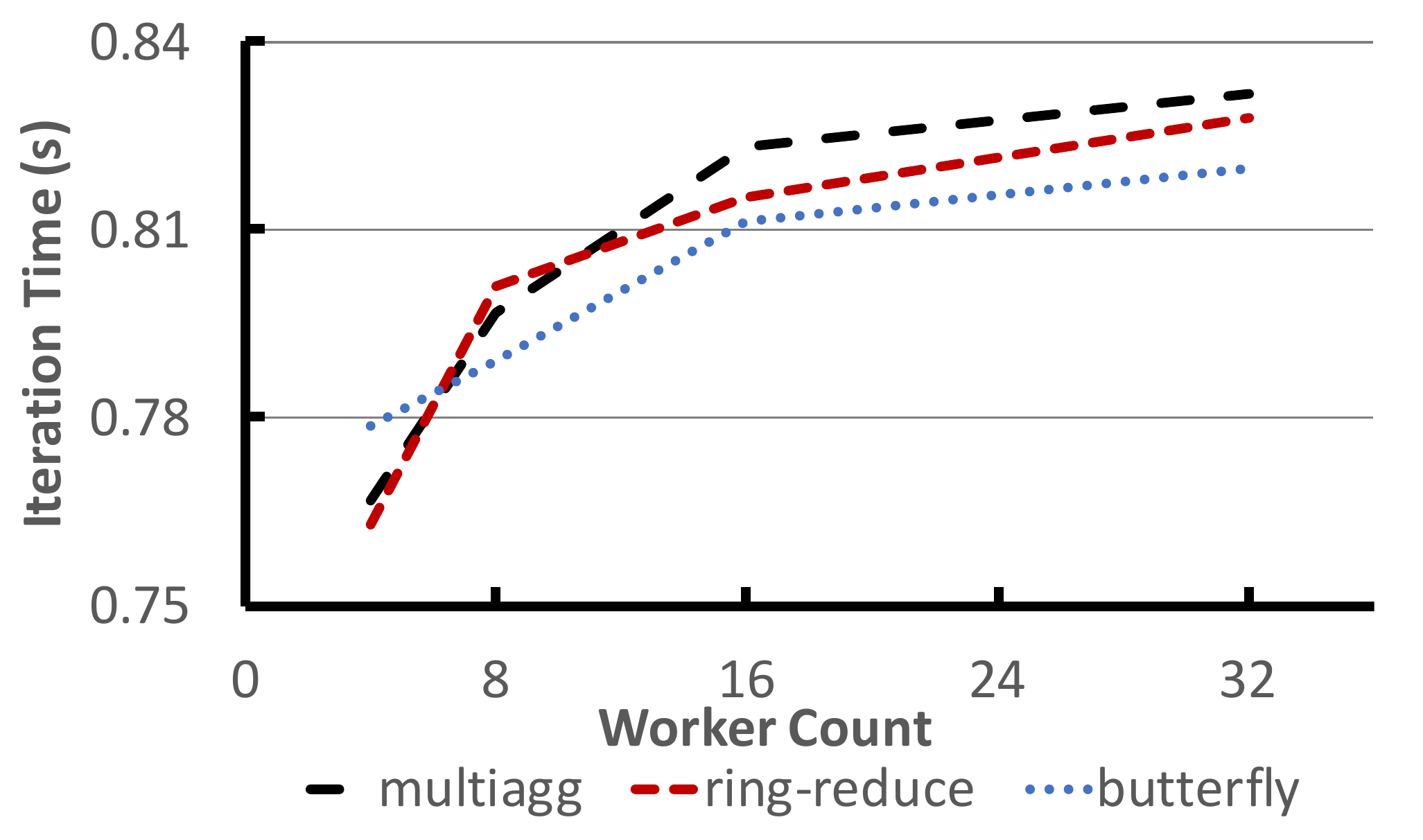}
    \caption{Varying Worker Count: Resnet-200 performance at 25~Gbps}
     \label{fig:resnet200_worker_ranking}
\end{minipage}
\hfill
\begin{minipage}{.3\textwidth}
    \includegraphics[width=\textwidth]{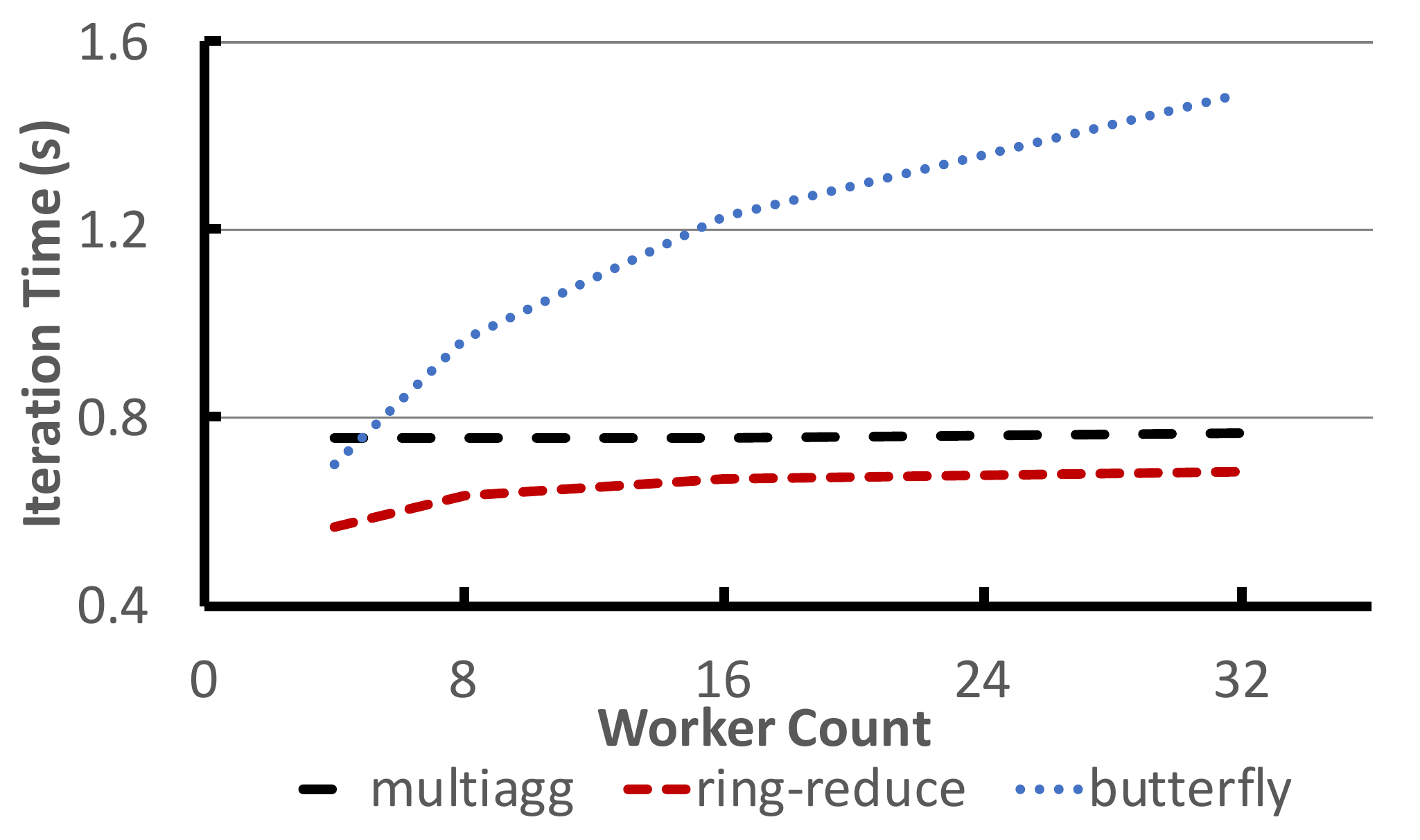}
    \caption{Varying Worker Count: VGG16 performance at 25~Gbps}
    \label{fig:vgg_worker_ranking}
\end{minipage}
\end{figure*}

\subsubsection{Butterfly Mixing vs. Ring-Reduce with messaging}
\label{sec:mixing_vs_messaging}
Butterfly mixing and ring-reduce are both impacted by an CNN model's backpropagation compute/network interactions, but in opposite ways. Butterfly mixing helps more for compute bound backpropagations, and ring-reduce helps more for more network bound backpropagation. In Table \ref{tab:all-reduce-speedup}, ring-reduce and butterfly mixing perform comparably for all models except VGG16, which has the most network bound backpropagation. This leads to stronger improvement from ring-reduce. Compare this to a lower bandwidth, 10~Gbps. Ring-reduce and butterfly mixing only perform comparably for Inception-v3, which has the most compute bound backpropagation. Conversely, ring-reduce outperforms butterfly mixing for Resnet-200 (Figure \ref{fig:resnet200_overall_ranking}), and VGG16 (Figure \ref{fig:vgg_overall_ranking}). 

\textit{Ring-reduce offers superior or equal performance impact as butterfly mixing.}

\subsection{Comparing End-host and In-Network Strategies}
\label{sec:endhost_vs_network}

In this section, we present overall simulations results for the top performing mechanisms seen so far: ring-reduce with messaging and multicast/in-network aggregation. We include results for butterfly mixing as another competitive reference point, but the focus of this section is on ring-reduce vs. multicast/in-network aggregation.

First, we fix the number of workers to 32 and vary bandwidth. The results for Inception-v3, Resnet-200, and VGG16 are shown in Figure~\ref{fig:inception_overall_ranking},  Figure~\ref{fig:resnet200_overall_ranking}, and Figure~\ref{fig:vgg_overall_ranking}, respectively. Next, for the same acceleration mechanisms, we show how performance scales with number of workers, while keeping bandwidth fixed at 25~Gbps. The results for Inception-v3, Resnet-200, and VGG16 are shown in Figure~\ref{fig:inception_worker_ranking}, Figure~\ref{fig:resnet200_worker_ranking}, and Figure~\ref{fig:vgg_worker_ranking}, respectively. Resnet-101 is left off this panel of graphs but shows consistent trends with Resnet-200.

These results indicate that with 32 workers -- across all bandwidths -- ring-reduce outperforms the combination of multicast and in-network aggregation. For Resnet-200, Resnet-101 and Inception-v3, ring-reduce and multicast+in-network aggregation perform very similarly. However, ring-reduce holds a key advantage in the VGG16 model. As shown in Figure \ref{fig:vgg_worker_ranking} ring-reduce has nearly a 2x performance advantage with 4 workers, and a 1.3x performance advantage with 32 workers. While the gap closes with more workers, we do not observe a case with VGG16 where multicast+in-network aggregation \textit{outperforms} ring-reduce. The gap between the two mechanisms is also more pronounced at low bandwidths, as seen in Figure \ref{fig:vgg_overall_ranking}.

Ring-reduce has one key advantage over multicast + in-network aggregation. The first and second ring of ring-reduce are equivalent to the parameter server's aggregation and distribution phase, respectively. While there is a per-worker local barrier between the aggregation and distribution phase, no such barrier exists for ring-reduce. The second \textit{distribution} ring of ring-reduce can proceed even while other parameters are still circulating their first ring. Thus the two phases can be pipelined, which enhances performance impact. The difference is particularly stark for models with disproportionately large penultimate layers, such as VGG16 in Figure \ref{fig:vgg_worker_ranking}. 

\textbf{Ring-reduce offers superior or equal performance impact as multicast with in-network aggregation.}

\subsection{All-reduce with network support}
\label{sec:all_reduce_netsupport}
\vspace{-.1in}
In this subsection, we explore the possibility of using in-network support to accelerate all-reduce algorithms. The primary combination we discuss in this section is ring-reduce with multicast. Note that multicast does not offer any performance benefits to butterfly mixing. The same applies for in-network aggregation for both ring-reduce and butterfly mixing. 
When using ring-reduce, multicast could be used to improve the performance time of the \textit{second ring}. When parameters have been distributed between workers equally, we find in our simulations that multicast in the second phase of ring-reduce has very limited impact on performance. This can be reasoned about analytically. The communication overhead in the second loop is: $\frac{Model\ size \times (Workers -1)}{Workers \times Bandwidth}$. When using multicast, the communication overhead is: $\frac{Model\ size}{Bandwidth}$ Note the bottleneck is on the receiving worker side. When the number of workers is large, these result in very similar performance. Our simulation results (Figure \ref{tab:all-reduce-speedup}) confirm that across the board, ring-reduce with multicast performs equivalently with just ring-reduce.

\subsection{Future (i.e., Synthetic) CNN Models}

\begin{figure*}
\begin{minipage}{.3\textwidth}
    \includegraphics[width=\textwidth]{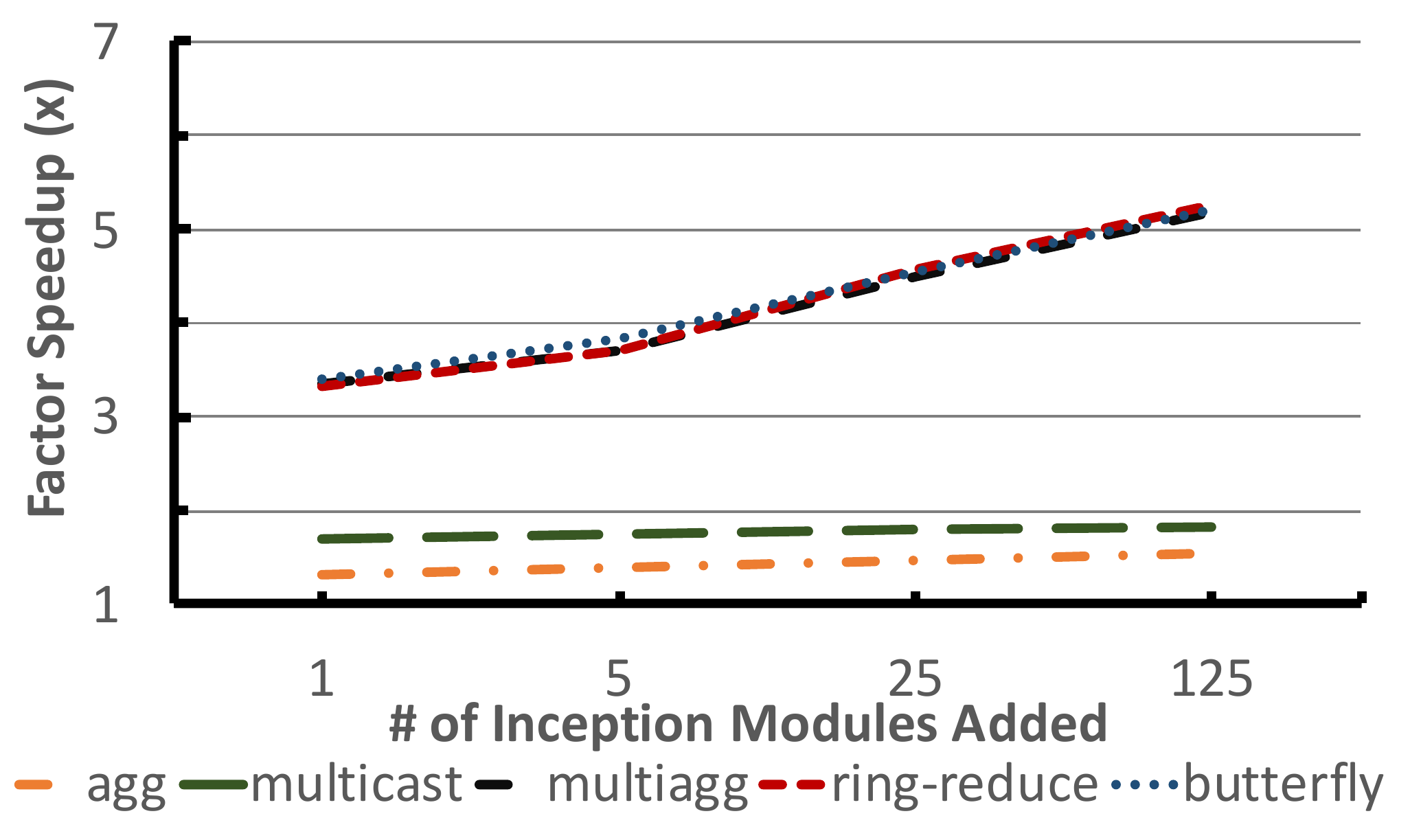}
    \caption{Synthetic Model (Network-heavy): Speedups relative to baseline increase as more network heavy layers are added}
    \label{fig:synthetic_network_heavy}
\end{minipage}
\hfill
\begin{minipage}{.3\textwidth}
    \includegraphics[width=\textwidth]{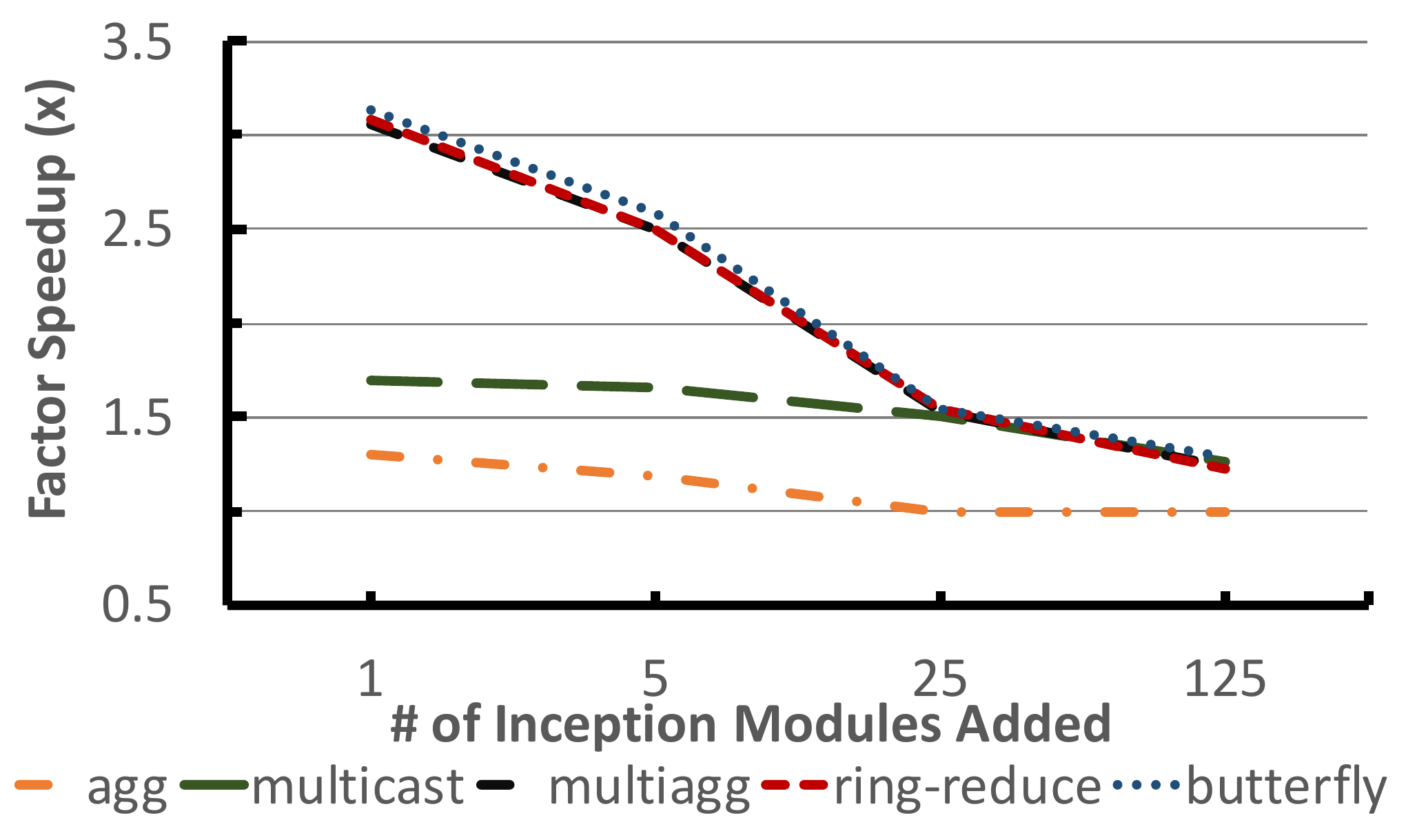}
    \caption{Synthetic Model (Compute-heavy): Speedups relative to baseline decrease as more compute heavy layers are added}
    \label{fig:synthetic_compute_heavy}
\end{minipage}
\hfill
\begin{minipage}{.3\textwidth}
    \includegraphics[width=\textwidth]{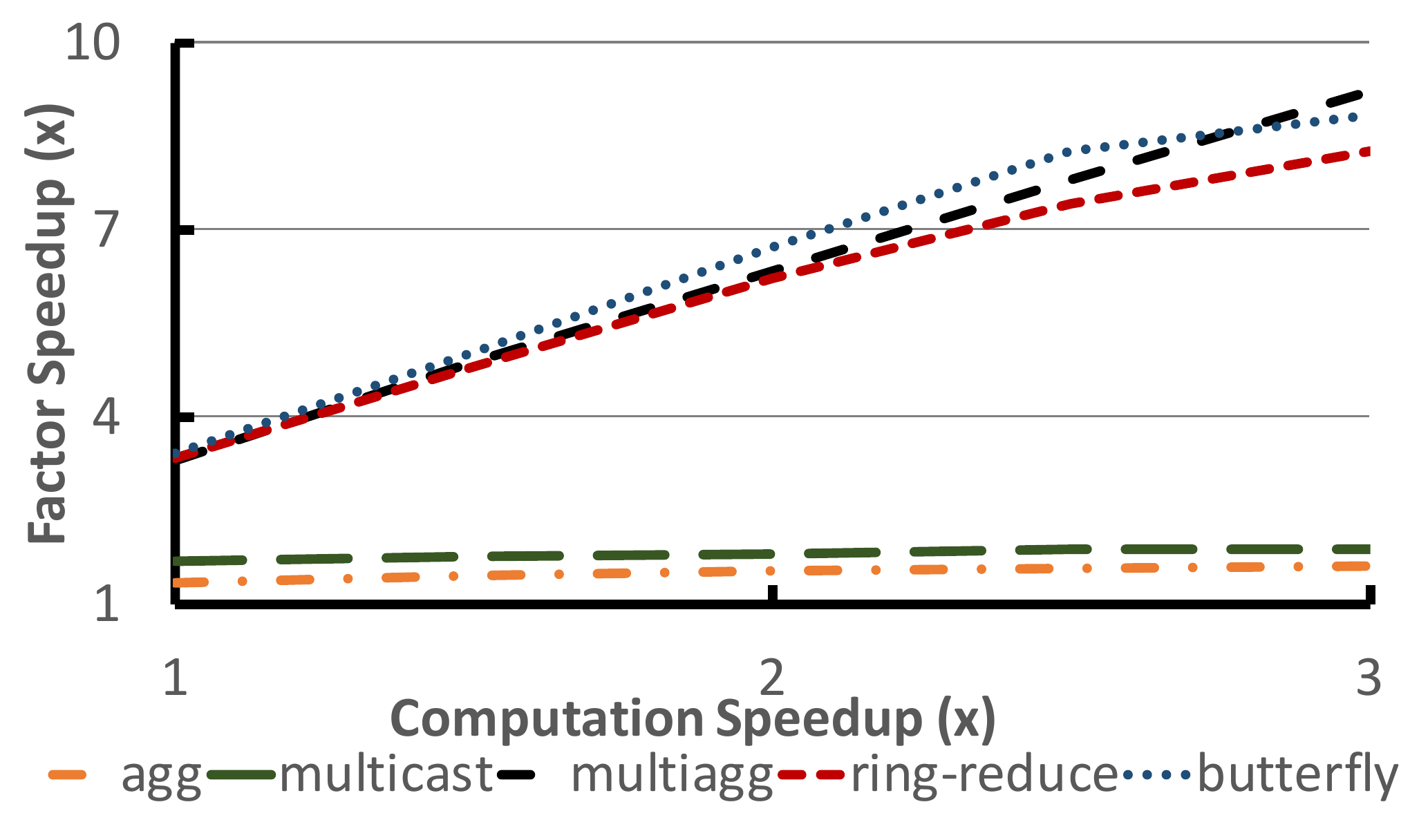}
    \caption{Faster GPU: Inception-v3 performance gains as compute speeds increase}
     \label{fig:inception_compute_speedup}
\end{minipage}
\end{figure*}

To this point in the paper, we have demonstrated that ring-reduce outperforms other proposals. Next we ask, how could these models change over time and would our results still hold? Recent model proposals have simply modified existing CNN models through the addition of convolutional layers~\cite{DBLP:journals/corr/HeZRS15, DBLP:journals/corr/HeZR016, 2018arXiv180805377E}. To test this, we add between 1 and 125 modules to the Inception-v3 model. We capture the full breadth of possible layers by adding one of two types of modules: Compute Intensive (35x35x288 module) and Network intensive (17x17x768 module). Again, we run our simulations with 32 workers with 25bps.
First, we observe in Figure \ref{fig:synthetic_compute_heavy}, \ref{fig:synthetic_network_heavy} that our mechanism ranking are preserved over both strains of synthetic models. However, the relative impact of the speedup mechanisms change. With the compute intensive synthetic model, the performance impact of in-network aggregation quickly drops to to zero, since pipelining in backpropagation provides ample time for parameters to be sent over the network between computations. In contrast, multicast provides the same amount of performance impact between layers, and ultimately equals the performance of the other mechanisms with only 25 layers added. If models trend towards becoming computationally expensive, \textbf{operators dedicated to using parameter servers could turn away from in-network aggregation for good, as multicast alone equals the impact of using both multicast and in-network aggregation}. For the network intensive synthetic model, multicast with in-network aggregation, ring-reduce, and butterfly mixing grow linearly with the number of layers. Even in the extreme case with 125 additional layers, neither in-network aggregation or multicast alone reaches it's maximal performance gain of 2x. 
While this is hardly an exhaustive list potential model changes, creating synthetic models from existing model traces is extremely simple. Operators and developers can easily modify existing traces to \textit{project} necessary changes to the network fabric or end-host design.
        
\subsection{Faster Computational Capabilities}
Next, we examine the case of potential enhancements in compute capabilities (e.g., faster hardware accelerators). In particular, do faster computations change the relative effectiveness of the acceleration mechanisms? In our examples, we increased the speed of convolutions by various factors. 

\begin{figure}
    \centering
    \includegraphics[width=0.44\textwidth]{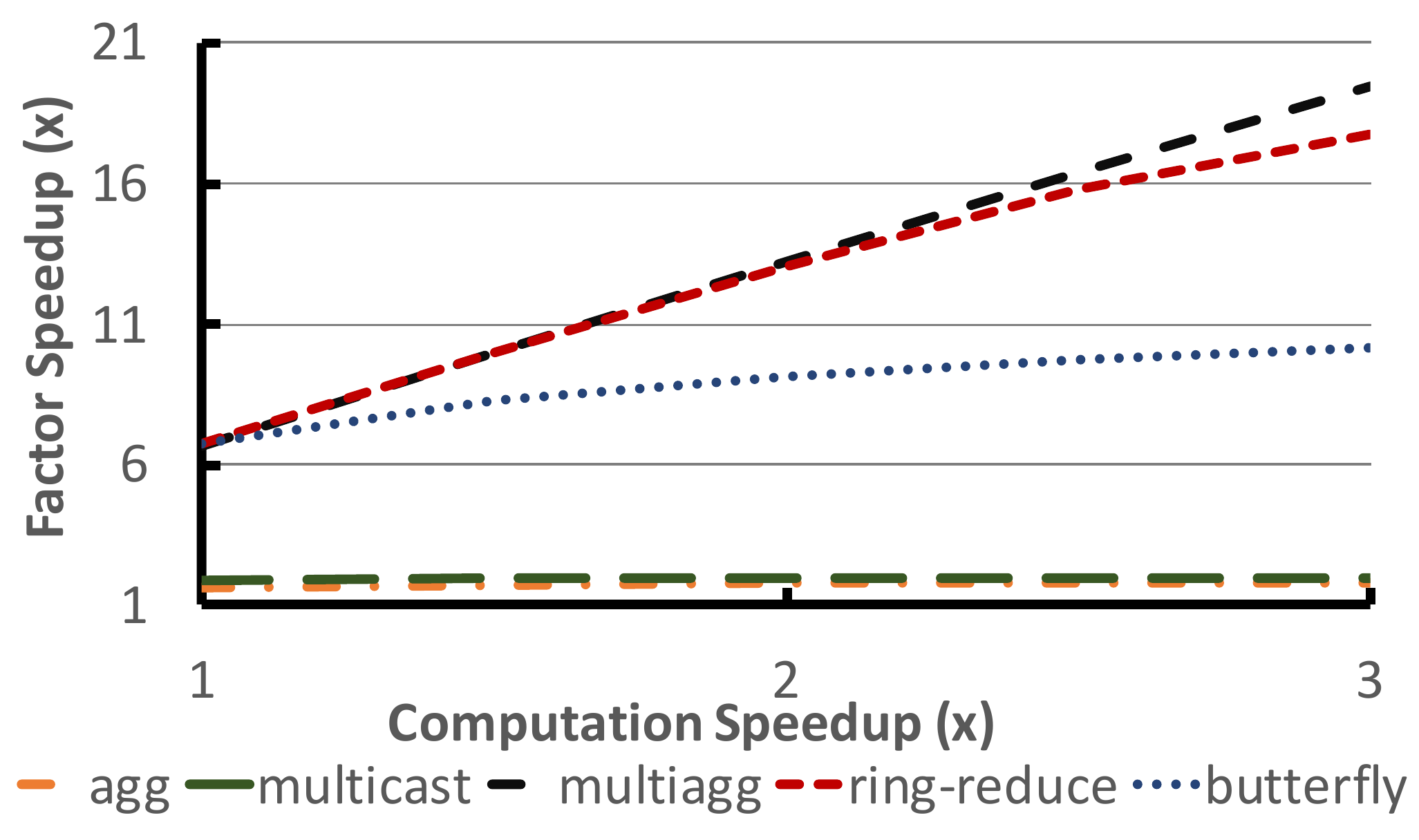}
    \caption{Faster GPU: Resnet-200 performance gains as compute speeds increase}
    \label{fig:resnet200_compute_speedup}
    \vspace{-.1in}
\end{figure}

Inception-v3 is shown in Figure \ref{fig:inception_compute_speedup} and Resnet-200 is shown in Figure \ref{fig:resnet200_compute_speedup}. First, we observe that there is a point where pipelined phases of the parameter server model become \textit{so network bound} that the parameter server paradigm (i.e., multicast with in-network aggregation) wins out. For most models, this point occurs at around 2.5x speedup, where multicast/in-network aggregation surpasses the performance of both ring-reduce and butterfly mixing. One standout trend from Figure \ref{fig:resnet200_compute_speedup} is in Resnet-200 at 3x computation speedup, where butterfly mixing results in only a 10x speedup while multicast with in-network aggregation leads to a 19.5x speedup. In fact, for all models except Inception-v3, butterfly mixing gains flags with faster computation. This is consistent with our observation that butterfly mixing impact decreases as backpropagation becomes more network bound. Inception-v3 happens to be so compute-bound that even at 3x compute speedup, butterfly mixing keeps pace with ring-reduce and multicast/in-network aggregation.

Our results indicate that faster compute capabilities could lead to better performance impact when jointly using multicast and in-network aggregation. We hasten to mention that performance is a function of many factors (e.g., bandwidth, memory, etc.) that grow at their own pace. The complexity of these assessment reinforces the need for a simple simulator that can be used to assess the current state of models and hardware.

\subsection{Evaluation Summary}
\label{sec:eval_summary}

From most to least performance impactful, our mechanism rankings are as follows: 1.) ring-reduce, 2.) multicast with in-network aggregation, 3.) butterfly mixing, 4.) multicast, 5.) in-network aggregation. Not only have we demonstrated that an \textbf{individual end-host mechanism outperforms joint usage of network support}, these performance rankings should continue to hold as models evolve and computation becomes increasingly accelerated. Ultimately, we hypothesize that the reason for this ranking is that current implementations of end-host mechanisms more effectively overlap distinct pipelined phases, thus exploiting available bandwidth more effectively over the entire iteration. This difference in pipelining efficiency grows more pronounced when the penultimate layers of the model are disporportionately large. As models inevitably change, operators must carefully examine the qualities of the final CNN layers when considering how to accelerate training.

\section{Robustness of Evaluation}
\label{sec:robustness}
Much like how we expect the CNN models to change, training software running on the end host will also change. During the simulator development, we identified several \emph{end host} design decisions that influence performance outcomes. Ultimately, we find that our findings are (mostly) robust to these end-host configurations. First, we look at \textit{equal} assignment of parameters to parameter servers. Next, we look at three design configurations that are not currently integrated into Tensorflow: Parameter Distribution, Message Pipelining, and removing parameter server side global barrier. In this section, we rely on simulations so to gain intuition about host level changes from a communication overhead point of view. Our simulations do not include system-level overheads for these approaches, and we leave evaluating this to future work.
    
\subsection{Parameter Assignment to Parameter Servers}
\label{sec:param_assignment}

\begin{table}[t]
\centering
\footnotesize
% \resizebox{\columnwidth}{!}{
\begin{tabular}{@{}lrrrr@{}}
\toprule
\multicolumn{1}{c}{\textbf{Num PS}} & \multicolumn{1}{c}{\textbf{CNN Model}} & \multicolumn{1}{c}{\textbf{Min \%}} & \multicolumn{1}{c}{\textbf{Max \%}} & \multicolumn{1}{c}{\textbf{Ideal \%}}\\ \midrule
4 & VGG-16              & $4 \cdot 10 ^{-5}$ & 0.918 & 0.25       \\
8 & VGG-16              & $1 \cdot 10 ^ {-5}$ &0.859 & 0.125     \\
4 & Inception-v3              & 0.0007          & 0.451  & 0.25       \\
8 & Inception-v3              & 0.0003          & 0.292  & 0.125       \\
4 & Resnet-200             & 0.165          & 0.33  & 0.25       \\
8 & Resnet-200             & 0.002          & 0.22  & 0.125       \\
 
\end{tabular}%}
\caption{Empirical Measurements for how parameter assignment to parameter servers. The columns show the percentage (in terms of bytes) of weights that are placed on the most and least occupied parameter server. Resnet-101 is similar to Resnet-200}
\label{table:param_distribution}
\vspace{-0.2in}
\end{table}

By default, TensorFlow iterates through the parameters in the model and assigns those parameters to the PS in a round robin fashion. While this effectively balances the \textit{number} of parameters per PS, the weights on each of them can be vastly different. The uneven distributions across several models are shown in Table \ref{table:param_distribution}. For example, in VGG-16, the fully connected layer alone consists of 5.44~Gb (out of 6.58~Gb total over the entire model). Through actual executions of TensorFlow, the simulations we have explored to this point accurately assigns parameters to parameter servers based on TensorFlow's default heuristic. Here, we explore this possibility of dividing parameters evenly. 

To simulate this fairly, we aggressively split each parameter between 8 parameter servers and 32 workers. Our results are shown in Table \ref{tab:multiple_ps}. With the exception of VGG16, ring-reduce continues to equal the combined efforts of multicast and in-network aggregation. If this end-host design change comes to fruition \textit{and} models trend towards VGG16 characteristics (i.e., short backpropagation and an exceptionally large last layer), operators should act accordingly to consider hardware network support. With the exception of VGG16, parameter assignments does not change the fact that end-host acceleration mechanisms outperform in-network acceleration.

\begin{table}[t]
\centering
\footnotesize
% \resizebox{\columnwidth}{!}{
\begin{tabular}{@{}lrrrr@{}}
\toprule
\multicolumn{1}{c}{\textbf{CNN Model}} &
\multicolumn{1}{c}{\textbf{Multiagg (s)}} & \multicolumn{1}{c}{\textbf{8 PS Multiagg (s)}} & \multicolumn{1}{c}{\textbf{Ring-Reduce (s)}}\\ \midrule
VGG-16 & 0.765 & 0.539 & 0.683 \\
Resnet-200 & 0.830 & 0.820 & 0.824 \\
Resnet-101 & 0.598 & 0.551 & 0.556 \\
Inception-v3 & 0.569 & 0.549 & 0.562
\end{tabular}%}
\caption{Using 8 parameter servers with a theoretically optimal distribution for multicast + aggregation does not provide substantive gains over ring-reduce. 32 workers, 25~Gbps}
\label{tab:multiple_ps}
\vspace{-0.2in}
\end{table}

\subsection{Message Pipelining}
\label{sec:message_pipelining}
In both the distribution and aggregation phases of distributed training, the node typically waits for the entirety of a parameter to arrive before sending it forward. This can be inefficient when individual parameters are large; for example the largest parameter in VGG16 model is in excess of 5~Gb. Instead, these parameters can be split up into smaller messages and forwarded when ready.

We incorporated message pipelining into our application and found, surprisingly, that for \textbf{all models we explored, communication within the parameter server model do not benefit whatsoever from message pipelining}. The improvements that result from message pipelining are swallowed by compute in the backpropagation. Only ring-reduce benefits significantly from messaging, thus our prior ring-reduce evaluatino in section \ref{sec:eval} keep equipped with messaging. Overall, the conclusions we presented in the evaluation are robust to message pipelining.

\subsection{Presence of a Global Barrier}
\label{sec:global_barrier}
Up to this point in the paper, we have used a global barrier in the parameter server. While this creates greater flexibility in operations that can be conducted over the entire model, this inhibits pipelining between iterations. Alternatively, this global barrier could be removed: when the parameter server receives all updates (from workers) to a parameter, it can immediately forward this update~\cite{DBLP:journals/corr/ZhangZXDHLHWXX17}. To fairly capture the performance change of removing the global barrier, we run three iterations and measure the time between a.) when the parameter server receives all updates from the first parameter during the latter part of the first iteration and b.) when that first parameter is received from all workers during the latter part of the third iteration. 

Table \ref{table:global_barrier} shows that the removal of the global barrier increases the impact of multicast plus in-network aggregation to the point that it becomes roughly equal to the impact of ring-reduce. The global barrier improves training time because the aggregation phase can be pipelined with the distribution phase. However, note that some of that improvement is returned because the worker cannot initiate forward pass until the first model layer arrives. That first layer is the final parameter computed in the backpropagation. 
Taking out the global barrier evens out the impact of ring-reduce and multicast/in-network aggregation, but our initial claim continues to hold.

\begin{table}[t]
\centering
\footnotesize
\begin{tabular}{@{}lrrrr@{}}
\toprule
\multicolumn{1}{c}{\textbf{CNN Model}} & \multicolumn{1}{c}{\textbf{Multiagg (s)}} & \multicolumn{1}{c}{\textbf{Ring-reduce (s)}} & \multicolumn{1}{c}{\textbf{Multiagg no barrier (s)}}\\ \midrule

VGG-16 & 1.53 & 1.37 & 1.76 \\
Resnet-200 & 1.65 & 1.65 & 1.65 \\
Resnet-101 & 1.17 & 1.13 & 1.08 \\
Inception-v3 & 1.14 & 1.13 & 0.988 \\
 
\end{tabular}%}
\caption{Removing the global barrier improves multicast + aggregation iteration time, but does not cause a decisive lead over ring-reduce. 32 workers, 25~Gbps}
\label{table:global_barrier}
\vspace{-0.2in}
\end{table}

\subsection{Parameter Distribution Order}
\label{sec:param_distribution_order}

In the parameter server model, there are two ways of distributing parameters: \textbf{round-robin distribution} (one model parameter at a time), and \textbf{block distribution} (send all model parameters in entirety to each worker, one at a time). While parameter distribution has been observed to be random~\cite{DBLP:journals/corr/abs-1803-03288}, simulator validation has shown that round-robin distribution order closely mimics actual empirical experiments. Surprisingly, in our experiments, we found that block distribution outperforms round-robin distribution.

When using round robin distribution, workers progress at roughly the same pace. Thus, until one of the workers begins the backpropagation process, the parameter server ingress bandwidth is un-utilized. When using block distribution, the parameter server bandwidth is utilized as soon as the first worker completes its forward pass. Moreover, only a single worker is likely to be doing backpropagation at a time, thus reducing incast.

How does block distribution compare to in-network aggregation? Recall that in-network aggregation benefits performance most when backpropagation staggering is minimal. When using block parameter distribution, under what analytical conditions would we \textit{similar performance impact between in-network aggregation and block distribution}? Let $B_1$ be the computation time of just the first layer of backpropagation, $B_N$ be the compute time of the entire back propagation, $N$ be the time to communicate the entire model, $Rem_{FP}$ be the remaining forward pass computation after the worker has received the entire model update from the parameter server. $B_1 + N + Rem_{FP} > Rem_{FP} + B_N$, which simplifies to $B_1 + N > B_N$. Thus, block distribution approximates the performance gains of in-network aggregation for models that have unusually large last layers and network transfer times (depends on model size and available network bandwidth).

Table \ref{tab:striping_raw} illustrates simulation results which show that block distribution performs similarly, or better, than in-network aggregation at vastly differently bandwidths. Block raises additional questions about the efficiency of using in-network aggregation for distributed CNN training.

\begin{table}[t]
    \centering
    \footnotesize
 \resizebox{\columnwidth}{!}{
    \begin{tabular}{@{}lrrrr@{}}
    \toprule
    \multirow{2}{*}{\textbf{CNN Name}} & \multirow{2}{*}{\textbf{Bandwidth (Gbps)}} & \multirow{2}{*}{\shortstack[r]{\textbf{Agg (s)} }} & \multirow{2}{*}{\textbf{Block Distr. (s)}}\\
    & & & & \\
    \midrule
        Inception-v3 & 10 & 2.99 & 3.1  \\
        VGG16 & 10 & 22.3 & 21.7 \\
        Resnet-101 & 10 & 4.9 & 4.94  \\
        Resnet-200 & 10 & 7.77 & 7.79 \\
        Inception-v3 & 100 & 0.71 & 0.77  \\
        VGG16 & 100 & 2.23 & 2.27 \\
        Resnet-101 & 100 & 0.89 & 0.94  \\
        Resnet-200 & 100 & 1.19 & 1.45 \\
    \end{tabular}}
    \caption{With 32 workers, the training time of using in-network aggregation and block (i.e., not round robin) parameter distributions is roughly the same}
    \label{tab:striping_raw}
    \vspace{-0.1in}
\end{table}

\section{Discussion}
\label{sec:discussion}

Next we briefly discuss the impact of other optimization strategies and systems considerations.

\noindent\textbf{Gradient Compression:} Other work~\cite{Lin2017DeepGC} has also looked at using gradient compression to reduce the amount of aggregation traffic sent during CNN training. Gradient compression and other compression techniques reduce model size but do not affect the number of network transfers. As a result, applying these methods is analogous to using a smaller CNN and is covered by our analysis.

\noindent\textbf{Asynchronous training:} Our analysis thus far has assumed the use of synchronous training algorithms (\S\ref{sec:background:ps_training}). We focused on these algorithms for ease of analysis and exposition, and our simulator and analysis techniques can be applied to asynchronous training algorithms. However, neither of the in-network mechanisms can be used for asynchronous training due to a lack of barriers between iterations.

\section{Conclusion}
We began this work wanting to develop network optimizations that improve CNN training performance. We found that despite a great deal of excitement about this area, little was understood about what types of optimizations were promising, or even how current optimizations impacted end-to-end CNN training performance. Thus we sought to address this question, and in doing so found that end host based solutions, which are arguably easier to deploy, generally provide better improvements than in-network solutions. We developed a trace-driven simulator, that simplifies the analysis of how network changes impact CNN performance. We hope that this simulator will provide a foundation to enable the community to develop and evaluate optimizations for improving CNN performance. We plan to open source the simulator and our data so as to allow the community to leverage and extend our findings.

\section{Acknowledgements}

We thank members of NetSys lab at UC Berkeley for their useful feedback, as well as Peter Gao and Qiyin Wu for their feedback on an early version of this work. This work was funded by NSF Grants 1817115, 1817116, 1704941, and was supported by Intel, VMware and Microsoft.

\bibliographystyle{abbrv}
\bibliography{ms}

\begin{thebibliography}{10}

\bibitem{Abadi2016TensorFlowAS}
M.~Abadi, P.~Barham, J.~Chen, Z.~Chen, A.~Davis, J.~Dean, M.~Devin,
  S.~Ghemawat, G.~Irving, M.~Isard, M.~Kudlur, J.~Levenberg, R.~Monga,
  S.~Moore, D.~G. Murray, B.~Steiner, P.~A. Tucker, V.~Vasudevan, P.~Warden,
  M.~Wicke, Y.~Yu, and X.~Zhang.
\newblock Tensorflow: A system for large-scale machine learning.
\newblock In {\em OSDI}, 2016.

\bibitem{Awan2018ScalableDD}
A.~A. Awan, J.~B{\'e}dorf, C.-H. Chu, H.~Subramoni, and D.~K. Panda.
\newblock Scalable distributed dnn training using tensorflow and cuda-aware
  mpi: Characterization, designs, and performance evaluation.
\newblock {\em CoRR}, abs/1810.11112, 2018.

\bibitem{baiduallred}
{Baidu Slicon Valley Lab}.
\newblock Bringing {HPC} techniques to deep learning.
\newblock
  \url{http://research.baidu.com/bringing-hpc-techniques-deep-learning/}.

\bibitem{barefoot}
{Barefoot Technology}.
\newblock \url{https://barefootnetworks.com/technology/}.

\bibitem{DBLP:conf/sdm/CannyZ13}
J.~F. Canny and H.~Zhao.
\newblock Butterfly mixing: Accelerating incremental-update algorithms on
  clusters.
\newblock In {\em Proceedings of the 13th {SIAM} International Conference on
  Data Mining, May 2-4, 2013. Austin, Texas, {USA.}}, pages 785--793, 2013.

\bibitem{Chen2000MPICO}
H.~A. Chen, Y.~O. Carrasco, and A.~W. Apon.
\newblock Mpi collective operations over ip multicast.
\newblock In {\em IPDPS Workshops}, 2000.

\bibitem{Chen2016RevisitingDS}
J.~Chen, R.~Monga, S.~Bengio, and R.~J{\'o}zefowicz.
\newblock Revisiting distributed synchronous sgd.
\newblock {\em CoRR}, abs/1604.00981, 2016.

\bibitem{Chen2015MXNetAF}
T.~Chen, M.~Li, Y.~Li, M.~Lin, N.~Wang, M.~Wang, T.~Xiao, B.~Xu, C.~Zhang, and
  Z.~Zhang.
\newblock Mxnet: A flexible and efficient machine learning library for
  heterogeneous distributed systems.
\newblock {\em CoRR}, abs/1512.01274, 2015.

\bibitem{cnnbenchmark}
{CNN Benchmarks}.
\newblock \url{https://github.com/jcjohnson/cnn-benchmarks#inception-v1}.

\bibitem{RFC1112}
S.~Deering.
\newblock Host extensions for {IP} multicasting.
\newblock STD~5, RFC Editor, August 1989.
\newblock \url{http://www.rfc-editor.org/rfc/rfc1112.txt}.

\bibitem{Dillon2017TensorFlowD}
J.~V. Dillon, I.~Langmore, D.~Tran, E.~Brevdo, S.~Vasudevan, D.~Moore,
  B.~Patton, A.~Alemi, M.~D. Hoffman, and R.~A. Saurous.
\newblock Tensorflow distributions.
\newblock {\em CoRR}, abs/1711.10604, 2017.

\bibitem{2018arXiv180805377E}
T.~{Elsken}, J.~{Hendrik Metzen}, and F.~{Hutter}.
\newblock {Neural Architecture Search: A Survey}.
\newblock {\em arXiv e-prints}, page arXiv:1808.05377, Aug. 2018.

\bibitem{Goyal2017AccurateLM}
P.~Goyal, P.~Doll{\'a}r, R.~B. Girshick, P.~Noordhuis, L.~Wesolowski,
  A.~Kyrola, A.~Tulloch, Y.~Jia, and K.~He.
\newblock {Accurate, Large Minibatch SGD: Training ImageNet in 1 Hour}.
\newblock {\em CoRR}, abs/1706.02677, 2017.

\bibitem{DBLP:journals/corr/abs-1803-03288}
S.~H. Hashemi, S.~A. Jyothi, and R.~H. Campbell.
\newblock Communication scheduling as a first-class citizen in distributed
  machine learning systems.
\newblock {\em CoRR}, abs/1803.03288, 2018.

\bibitem{DBLP:journals/corr/HeZRS15}
K.~He, X.~Zhang, S.~Ren, and J.~Sun.
\newblock {Deep Residual Learning for Image Recognition}.
\newblock {\em CoRR}, abs/1512.03385, 2015.

\bibitem{DBLP:journals/corr/HeZR016}
K.~He, X.~Zhang, S.~Ren, and J.~Sun.
\newblock {Identity Mappings in Deep Residual Networks}.
\newblock {\em CoRR}, abs/1603.05027, 2016.

\bibitem{Hoefler2007APC}
T.~Hoefler, C.~Siebert, and W.~Rehm.
\newblock A practically constant-time mpi broadcast algorithm for large-scale
  infiniband clusters with multicast.
\newblock {\em 2007 IEEE International Parallel and Distributed Processing
  Symposium}, pages 1--8, 2007.

\bibitem{8241753}
J.~Ker, L.~Wang, J.~Rao, and T.~Lim.
\newblock Deep learning applications in medical image analysis.
\newblock {\em IEEE Access}, 6:9375--9389, 2018.

\bibitem{Li2014ScalingDM}
M.~Li, D.~G. Andersen, J.~W. Park, A.~J. Smola, A.~Ahmed, V.~Josifovski,
  J.~Long, E.~J. Shekita, and B.-Y. Su.
\newblock {Scaling Distributed Machine Learning with the Parameter Server}.
\newblock In {\em OSDI}, 2014.

\bibitem{Lin2017DeepGC}
Y.~Lin, S.~Han, H.~Mao, Y.~Wang, and W.~J. Dally.
\newblock Deep gradient compression: Reducing the communication bandwidth for
  distributed training.
\newblock {\em CoRR}, abs/1712.01887, 2017.

\bibitem{LITJENS201760}
G.~Litjens, T.~Kooi, B.~E. Bejnordi, A.~A.~A. Setio, F.~Ciompi, M.~Ghafoorian,
  J.~A. van~der Laak, B.~van Ginneken, and C.~I. Sánchez.
\newblock A survey on deep learning in medical image analysis.
\newblock {\em Medical Image Analysis}, 42:60 -- 88, 2017.

\bibitem{LuoKrishnamurthy}
L.~Luo, M.~Liu, J.~Nelson, L.~Ceze, A.~Phanishayee, and A.~Krishnamurthy.
\newblock {Motivating In-network Aggregation for Distributed Deep Neural
  Network Training}.
\newblock In {\em Workshop on Approximate Computing Across the Stack}. ACM,
  2017.

\bibitem{Mai2015OptimizingNP}
L.~Mai, C.~Hong, and P.~Costa.
\newblock {Optimizing Network Performance in Distributed Machine Learning}.
\newblock In {\em HotCloud}, 2015.

\bibitem{10.1007/978-3-540-24685-5_1}
R.~Rabenseifner.
\newblock Optimization of collective reduction operations.
\newblock In M.~Bubak, G.~D. van Albada, P.~M.~A. Sloot, and J.~Dongarra,
  editors, {\em Computational Science - ICCS 2004}, pages 1--9, Berlin,
  Heidelberg, 2004. Springer Berlin Heidelberg.

\bibitem{Sapio2017InNetworkCI}
A.~Sapio, I.~Abdelaziz, A.~Aldilaijan, M.~Canini, and P.~Kalnis.
\newblock In-network computation is a dumb idea whose time has come.
\newblock In {\em HotNets}, 2017.

\bibitem{DBLP:journals/corr/SimonyanZ14a}
K.~Simonyan and A.~Zisserman.
\newblock {Very Deep Convolutional Networks for Large-Scale Image Recognition}.
\newblock {\em CoRR}, abs/1409.1556, 2014.

\bibitem{Tomov2018OnDN}
N.-S. Tomov and S.~Tomov.
\newblock On deep neural networks for detecting heart disease.
\newblock {\em CoRR}, abs/1808.07168, 2018.

\bibitem{horovod}
{Uber}.
\newblock {Meet Horovod: Uber's Open Source Distributed Deep Learning Framework
  for TensorFlow}.
\newblock \url{https://eng.uber.com/horovod/}.

\bibitem{Vishnu2016DistributedTW}
A.~Vishnu, C.~Siegel, and J.~Daily.
\newblock Distributed tensorflow with mpi.
\newblock {\em CoRR}, abs/1603.02339, 2016.

\bibitem{DBLP:journals/corr/WallachDH15}
I.~Wallach, M.~Dzamba, and A.~Heifets.
\newblock Atomnet: {A} deep convolutional neural network for bioactivity
  prediction in structure-based drug discovery.
\newblock {\em CoRR}, abs/1510.02855, 2015.

\bibitem{Yuan2002GroupMS}
X.~Yuan, S.~Daniels, A.~Faraj, and A.~Karwande.
\newblock Group management schemes for implementing mpi collective
  communication over ip-multicast.
\newblock In {\em JCIS}, 2002.

\bibitem{DBLP:journals/corr/ZhangZXDHLHWXX17}
H.~Zhang, Z.~Zheng, S.~Xu, W.~Dai, Q.~Ho, X.~Liang, Z.~Hu, J.~Wei, P.~Xie, and
  E.~P. Xing.
\newblock Poseidon: An efficient communication architecture for distributed
  deep learning on {GPU} clusters.
\newblock {\em CoRR}, abs/1706.03292, 2017.

\end{thebibliography}
\end{document}